\documentclass[showpacs, showkeys, twocolumn, amsmath,amssymb]{revtex4}
\usepackage[utf8]{inputenc}
\usepackage{ulem}
\usepackage{mathrsfs}
\usepackage{mathbbold}
\usepackage{indentfirst}
\usepackage{amsfonts}
\usepackage{amssymb}%
\usepackage{color}
\usepackage{slashed}
\usepackage{graphicx}
\usepackage{extarrows}

\def\al{\alpha}
\def\be{\beta}
\def\ga{\gamma}

\def\th{\theta}

\def\ka{\kappa}
\def\la{\lambda}

\def\si{\sigma}

\def\ph{\phi}

\def\Ga{\Gamma}

\def\Th{\Theta}

\def\mn{{\mu\nu}}

\def\prt{\partial}

\newcommand{\beq}{\begin{equation}}
\newcommand{\eeq}{\end{equation}}
\newcommand{\bea}{\begin{eqnarray}}
\newcommand{\eea}{\end{eqnarray}}
\newcommand{\bseq}{\begin{subequations}}
\newcommand{\eseq}{\end{subequations}}

\newcommand{\nn}{\nonumber\\}

\def\lsim{\mathrel{\rlap{\lower4pt\hbox{\hskip1pt$\sim$}}
    \raise1pt\hbox{$<$}}}
\def\gsim{\mathrel{\rlap{\lower4pt\hbox{\hskip1pt$\sim$}}
    \raise1pt\hbox{$>$}}}

\def\prt{\partial}

\def\mn{{\mu\nu}}

\newcommand{\ie}{{\it i.e.}}
\newcommand{\etc}{{\it etc}}
\newcommand{\eg}{{\it e.g.}}

\newcommand{\hf}{\frac{1}{2}}
\newcommand{\bit}{\begin{itemize}}
\newcommand{\eit}{\end{itemize}}

\providecommand{\Journal}[4] {#1 {\bf#2}, #4 (#3)}
\providecommand{\CQG}{Class. Quantum Grav.}%
\providecommand{\CMP}{Commun. Math. Phys.}%
\providecommand{\LRR}{Living Rev. Relativ.}%
\providecommand{\PR}{Phys. Rev.} %
\providecommand{\PRL}{Phys. Rev. Lett.} %
\providecommand{\PRD}{Phys. Rev. D} %
\providecommand{\PRSLA}{Proc. R. Soc. Lond. A} %
\providecommand{\RMP}{Rev. Mod. Phys.} %
\providecommand{\PLB}{Phys. Lett. B} %
\providecommand{\pOE}{Opt. Express} %

\providecommand{\NT}{Nature} %
\providecommand{\NPB}{Nucl. Phys. B} %

\providecommand{\AnPhys}{Ann. Phys.} %
\providecommand{\APJ}{ApJ} %
\providecommand{\JMP}{J. Math. Phys.} %
\providecommand{\JHEP}{JHEP} %
\providecommand{\JCAP}{JCAP} %

\begin{document}
\title{\Large {The asymptotic behavior of Lorentz-violating photon fields}}

\author{Zhi Xiao${}^{a, b}$}
\email{spacecraft@pku.edu.cn}
\author{Hao Wang${}^{a, b}$}
\affiliation{${}^{a}$North China Electric Power University, Beijing 102206, China}
\affiliation{${}^{b}$Hebei Key Laboratory of Physics and Energy Technology, North China Electric Power University, Baoding 071000, China}

\begin{abstract}
In this work, we derive the Newman-Penrose formalism of Maxwell's equations using two approaches:
differential forms and intrinsic derivatives.
Denoting $(k_{AF})^\mu$ as $k^\mu$, with $k^\mu=(k^t,k^r,0,0)$ in spherically symmetric spacetimes,
we show that the expansion in $r^{-1}$ fails to produce consistent, closed solutions due to the inability to separate Lorentz-violating (LV) phase factors,
as the Lorentz-invariant (LI) null tetrad does not adapt to the LV wavefront.
Moreover, with exact formal solutions, we demonstrate that the expansion is nonperturbative in the LV parameter $k^2\equiv k^t-k^r$.
For $r\gg1/k^2$, higher powers of $k^2$ dominate over lower powers, as the latter decay more rapidly with increasing $r$.
Although the Coulomb mode $\phi_1\sim\mathcal{O}(\ln{r}/r^2)$ deviates from the LI expectation $\mathcal{O}(r^{-2})$
due to LV corrections, the leading outgoing radiation mode remains unaffected, i.e., $\phi_2\sim\mathcal{O}(r^{-1})$.
Given the constraint $|k_{AF}|\le10^{-44}$GeV \cite{CMBLV-N}, the three complex scalars $\phi_a$ ($a=0,1,2$)
still obey the peeling theorem: $\phi_a\sim\mathcal{O}(r^{(a-3)}),~a=0,1,2$ for large, finite distances.
\end{abstract}

\providecommand{\keywords}[1]
{
  \small	
  \textbf{\textit{Keywords---}} #1
}



\maketitle
\section{Introduction}
Lorentz symmetry (LS) is a fundamental spacetime symmetry in both general relativity (GR) and quantum field theory (QFT) in particle physics.
Unlike various internal symmetries, Lorentz symmetry plays a crucial role in the formulation of relativistic QFT and GR \cite{WeinbergQFT1, WeinbergGrav}.
However, recent developments in searching for a satisfactory quantum theory of gravity have suggested that Lorentz symmetry may not be an exact symmetry \cite{LVMotivations}
and could even be an emergent symmetry at low energies \cite{EmergentS}.
Such a possibility has spurred numerous experimental tests and also led to the development of a broad framework in the spirit of effective field theory (EFT)
attempting to incorporate various kinds of Lorentz symmetry violation (LSV), the Standard Model Extension (SME) \cite{SMEa,SMEb,SMEg}.
The SME has significantly advanced experimental efforts to constrain LS through particle physics and astronomical observations \cite{dataTable}.
In addition to the SME, alternative approaches to describing LSV have emerged that go beyond EFT, such as very special relativity \cite{VSR2006} and doubly special relativity \cite{DSR}.

Interestingly, Lorentz symmetry itself originates from the study of Maxwell's electrodynamics, which,
in fact, possesses a much larger symmetry group, the conformal symmetry group.
In this context, we expect the study of electrodynamics in the SME framework may reveal new features of the Lorentz violating (LV) effects.
Actually, the LSV in electrodynamics has been generalized to arbitrary dimensional operators
and nonabelian gauge groups by Kostleck\'y, Mewes and Zonghao Li \cite{SMEb}.
This work primarily focus on the power-counting renormalizable photon operators,
namely, the $k_F$ \cite{Marco2012} and $k_{AF}$ terms in the minimal SME (mSME) \cite{SMEa}, which has been thoroughly studied.
However, to date, there has been little attention paid to the study of the asymptotic behaviors of LV electromagnetic fields at large distances.
An exception is the study of the consistency of spontaneous LSV with asymptotic flatness, assuming all fields are static and spherical symmetric \cite{Yuri2021Asym},
where it was found that the Weyl-like ``t" term cannot allow asymptotic flatness solutions, thereby resolving the ``t-puzzle".
Our goal is more moderate, instead of tackling the asymptotic behaviors of LV gravity at large distance,
we aim to explore the asymptotic behavior of LV electrodynamics, which is a much simpler system than gravity.
This may help to bridge the gap in understanding the asymptotic behaviors of LV-modified long range forces,
with an emphasis on the infrared rather than ultraviolet behaviors.

It is important to note that {\it the asymptotic behavior in this context is distinct from the ``asymptotic states"}
discussed in QFT, though there are some similarities.
In QFT, asymptotic states refer to the initial and final states of a system as it approaches infinity past
or future. In these cases, the states are effectively described as free particle states,
since collisions occur within a microscopically small region and over a brief period.
As a result, there is essentially no interaction between the incoming or outgoing beams of particles long before
or long after the collision.
Asymptotic states play a crucial role in Lehmann-Symanzik-Zimmermann reduction formula
and tests of LS often involve kinetics of extern states,
making them a central topic in the literature \cite{ASQFT2014,ASpin0-2017}.
However, asymptotic behavior describes how fields decay as the distance from the source increases, particularly at spatial or null infinities.
In this sense, asymptotic behavior stresses more on the classical aspects of fields, as any quantum effect must be averaged over large distances.
In short, the study of asymptotic behavior aims to isolate source from the rest of the world and emphasizes on long-range behaviors,
while the study of asymptotic states aims to isolate interactions within a localized region of spacetime to allow well-defined initial or final states of quantum particles.

The advantages of studying the asymptotic behavior of electromagnetic fields at large distances include:
1. it can clearly disentangle radiation modes from Coulomb or other modes.
2. massless field equations are conformal invariant, enabling conformal transformation to compactify null infinity into a finite region \cite{Penrose65},
simplifying the analysis of radiation patterns and energy flux at null infinity.
3. in LI theory, massless particles travel along null geodesics.
So the null formalism is particularly suitable for studying the properties of massless fields.
Recent studies reveal that it effectively isolates the physical degree of freedoms responsible for the infrared structure of massless fields at large distances.
This can help to uncover the deep connection between large gauge transformations and asymptotic symmetries,
as highlighted by the Weinberg soft theorem \cite{1965Weinberg} and electromagnetic memories \cite{EM-Memory17}.
We wonder whether these results still hold in modified electrodynamics,
such as the LV electrodynamics, and focus primarily on the peeling-off theorem, which describes {\it the falloff
of the electromagnetic fields at null infinity in powers of $1/r$}.
In this context, strictly speaking, the vacuum is no longer the electromagnetic vacuo,
but rather one filled with various LV background fields.

To study LV electrodynamics with the null formalism, we must consider the light cone structure,
which is typically altered in most LV scenarios.
However, it seems at least at leading order in the LV coefficients,
the null formalism may still be applied to examine the asymptotic properties of photon fields.
After briefly review the photon sector in the minimal SME in subsection \ref{reviewFlat},
we investigate the parallel propagation of the polarization vector and photon flux along the wave vector
in the short wavelength approximation in subsection \ref{curveMaxw}.
In Sec. \ref{NullTetrad}, we review the basics of null formalism
and derive the Newman-Penrose (NP) form of Maxwell equations for both LI and LV theories.
In Sec. \ref{FormalSol}, we focus on the formal solution of the CPT-odd NP Maxwell equations,
showing that the equations adapted to the LI null tetrad cannot yield consistent and closed
asymptotic expansions in the affine parameter $r$, due to the failure to separate the LV phase factors.
Further analysis of the exact integral reveals that the expansion may be nonperturbative in small LV parameters,
supporting the nonpeturbative polarization structure of CPT-odd Maxwell theory discovered in Ref. \cite{vacuoCerenRad}.
In Sec. \ref{EMTNull}, we briefly analyze the energy momentum tensor within the NP formalism,
which provides relatively loose but useful constraints on the falloff behavior of Maxwell fields,
partially confirming our earlier results.
Finally, we conclude with a brief summary in Sec. \ref{Summary}.
In this paper, the metric signature is $(-,+,+,+)$ and the convention for the totally antisymmetric Levi-civita tensor is $\epsilon_{0123}=+1$.
The greek indices $\mu,~\nu,~\rho,...$ are for spacetime manifold and the Latin indices $a,b,c,...$ are for frame or tetrad.

\section{The Lorentz-violating extension of Maxwell equations}\label{miniSME-ga}
The action of the photon field in the mSME is constructed from the photon field $A_\mu$ and  background tensor fields $(k_{AF})^\mu$ and $(k_F)^{\mn\rho\si}$, and is given by
{\small
\bea\label{Maxwell-I}&&
\hspace{-6mm}
I_A=-\int\frac{\sqrt{-g}}{4}d^4x\left[\chi^{\mn\rho\si}F_{\mn}F_{\rho\si}-4(k_{AF})^\mu(^*F_{\mn})A^\nu\right],
\eea
}\hspace{-.25mm}where $\chi^{\mn\rho\si}\equiv\,g^{\mu[\rho}g^{\si]\nu}+(k_F)^{\mn\rho\si}$ and ${^*F}_{\mn}\equiv\hf\epsilon_{\mn\rho\si}F^{\rho\si}$.
The Lagrangian in (\ref{Maxwell-I}) is power counting renormalizable.
As for LV photon operators with dimension higher than $4$, interested reader may refer to
Ref. \cite{SMEb}, where the latter two articles specifically address higher dimensional operators up to arbitrary dimension,
including terms that describe effective photon self-interactions.
The equation of motion for the action $I=I_A+\int\sqrt{-g}\,d^4x\,j\cdot A$ is
\bea\label{Maxwell-LVG}&&
\hspace{-3mm}\nabla_\mu[F^{\mn}+(k_F)^{\mn\rho\si}F_{\rho\si}]+\epsilon^{\mn\rho\si}(k_{AF})_\mu F_{\rho\si}\nn&&~~~
-\epsilon^{\mn\rho\si}A_\si\nabla_\mu[(k_{AF})_\rho]=-j^\nu,
\eea
where $j^\nu\equiv\bar{\psi}\Ga^\mu\psi$.  
To investigate the LV effects of the photon field in curved spacetime,
we'd better first revisit some known properties of the LV corrected Maxwell equations in flat spacetime.

\subsection{Maxwell equations in Minkowski spacetime}\label{reviewFlat}
There are two distinct LV coefficients, each exhibiting different behaviors under CPT transformation.
To gain a clearer physical understanding, it is better to treat them separately.

\subsubsection{CPT-odd $(k_{AF})^\mu$ backgrounds}
To simplify our notation, we denote $k$ instead of $k_{AF}$ in the following.
In the flat spacetime with $k_F=0$, Eq. (\ref{Maxwell-LVG}) becomes
\bea\label{InhomoMaxw}&&
\nabla\cdot\vec{E}=\rho+2\vec{k}\cdot\vec{B},~~
\nabla\times\vec{B}-\dot{\vec{E}}=\vec{j}+2k^0\vec{B}-2\vec{k}\times\vec{E}.\nn
\eea
The homogeneous Maxwell equations
\bea
\nabla\cdot\vec{B}=0,\quad
\nabla\times\vec{E}+\dot{\vec{B}}=0
\eea
remain unaltered, because $dF=0$ comes from the Bianchi identity, which indicates $F=dA$.
The dispersion relation corresponding to Eq. (\ref{InhomoMaxw}) reads
\bea\label{QuarticCPTO}
(p^2)^2+4p^2k^2-4(p\cdot k)^2=0.
\eea
This is a quartic equation and has 4 solutions in general.
The two among them are negative energy solutions,
and the other two different branches correspond to birefringence solutions.
In general, the analytical expressions are rather involved and a closed solution cannot be obtained,
however, an implicit analytic solution
\bea
\omega_{\pm}^2-\vec{p}^2=\mp2[k^0\vec{p}-\vec{k}\cdot\frac{\vec{p}}{|\vec{p}|}\omega_{\pm}]
(1-4\frac{(\vec{k}\times\vec{p})^2}{\omega_{\pm}^2-\vec{p}^2})^{-\hf},
\eea
was given in Ref. \cite{CFJ1990}, where the theory was first proposed.

Assuming the observer background vector $k^\mu$ is timelike $k^2<0$,
we can always transform to the particular frame with $k^\mu=(k^0,\vec{0})$.
Similarly for spacelike $k^2>0$, $k^\mu=(0,\vec{k})$, and for lightlike $k^2=0$,
$k^\mu=(|\vec{k}|,\vec{k})$ in their specific frames.
Defining $\bar{p}=\sqrt{\vec{p}^2}$, the corresponding solutions of Eq. (\ref{QuarticCPTO}) are given by
\bseq\label{CPTV-DRLC}
\bea&&
\hspace{-9.6mm}|\omega|=\bar{p}\sqrt{1\pm\frac{2k^0}{\bar{p}}},\qquad\qquad\qquad~~~~~k^\mu=(k^0,\vec{0});\\&&
\hspace{-9.6mm}|\omega|=\left[{\bar{p}^2+{2\vec{k}^2}\pm2\sqrt{({\vec{k}^2})^2
+{(\vec{k}\cdot\vec{p})^2}}}\right]^{\hf},
k^\mu=(0,\vec{k});\\&&
\hspace{-9.6mm}|\omega|=\bar{p}\sqrt{1+\frac{\vec{k}^2}{\bar{p}^2}\mp\frac{2\vec{k}\cdot\vec{p}}{\bar{p}^2}}
\pm|\vec{k}|,
\quad~~ k^\mu=(|\vec{k}|,\vec{k}).
\eea
\eseq
where we use the absolute value $|\omega|$, to avoid the distinction between negative energy solutions from positive ones.
The $\pm$ signs in the square root in the above dispersion relation clearly show that
there are two branches of polarization modes, of which the propagation velocities are quite different
as long as the magnitudes of the LV coefficients are large enough.
Moreover, there could be tachyonic instabilities in the ultra-long wavelength region, where $2|k^0|>\bar{p}$ for timelike $k^\mu$ and $|\omega|$ becomes imaginary.
As for lightlike and spacelike $k^\mu$, there are no such tachyonic issues.

The stress-energy tensor for the action (\ref{Maxwell-I}) with $k_F=0$ in flat spacetime is given by
\bea\label{OgaEMT}&&
\theta_O^{\mn}=\theta_0^{\mn}+k^\nu{^*F}^{\mu\be}A_{\be},
\eea
where $\theta_0^{\mn}\equiv F^{\mu\al}F^\nu_{~\al}-\frac{\eta^{\mn}}{4}F_{\al\be}F^{\al\be}$ is the stress energy tensor corresponding to the conventional LI Maxwell theory.
One can clearly see that the LV contribution to $\th^{\mn}$ is gauge dependent. A naive analysis shows that
$-k^0\vec{B}\cdot\vec{A}$ may induce instabilities due to the non-positive
definiteness of the energy density. 
However, a detailed analysis shows the total energy-momentum tensor, obtained from an integral over all space,
is gauge invariant, since the gauge-dependent term from the gauge transformation is a pure surface term \cite{CFJ1990}.
Furthermore, the negative energy flux, which could raise instability issues in the long wavelength region where $2|k^0|>\bar{p}$,
is shown to be crucial for balancing the net positive energy radiated away by the vacuum Cerenkov radiation \cite{vacuoCerenRad,CFJ1990}.
The naive condition for the occurrence of this radiation is when the phase and group velocities $v_p=v_g=\frac{1}{\sqrt{1\pm\frac{2k^0}{\bar{p}}}}\sim1\mp\frac{2k^0}{\bar{p}}$, are less than 1,
which exactly occurs when $2|k^0|>\bar{p}$, even for arbitrarily small $\bar{p}$.

\subsubsection{CPT-even $(k_{F})^{\mn\rho\si}$ backgrounds}
The CPT-even coefficient $(k_{F})^{\mn\rho\si}$ is more complicated.
Due to its symmetry, which resembles that the Riemann tensor, in general it contains 19 parameters,
As a result, it can be decomposed into the birefringent Weyl-like part and the birefringence-free (in linear order) \cite{Marco2015, Cosbire2011} Ricci-like part,
\bea\label{kFdecom}
(k_F)_{\mn\rho\si}=(W_F)_{\mn\rho\si}+\frac{2}{N-2}[g_{\mu[\rho}(c_F)_{\si]\nu}
-g_{\nu[\rho}(c_F)_{\si]\mu}].
\eea
If we set $k_{AF}=0$ and the Weyl-like part $(W_F)_{\mn\rho\si}=0$ in $k_F$,
we obtain the modified Maxwell equations given below
\bea&&
\hspace{-3.6mm}\left[1-\left(c_{F}\right)_{00}\right]\nabla\cdot{\vec{E}}+\left(c_{F}\right)_{i j}\partial_{i} E^{j}-\left(c_{F}\right)_{0 i}(\nabla\times{{\vec{B}}})^{i}=\rho,\nn&&
\hspace{-3.6mm}(\nabla\times{{\vec{B}}})^{i}-\left[1-\left(c_{F}\right)_{00}\right] \dot{E}^{i}+\left(c_{F}\right)_{i j}(\nabla\times{{\vec{B}}}-\dot{\vec{E}})^{j} \nn&&
+\left[\left(c_{F}\right)_{0 i} \nabla\cdot{{\vec{E}}}-\left(c_{F}\right)_{0 j} \partial_{j} E^{i}\right]\nn&&
+\epsilon_{ijk}\left[\left(c_{F}\right)_{jl} \partial_{l} B^{k}-\left(c_{F}\right)_{0 j}\dot{B}^{k}\right]=j^i.
\eea
The Weyl-like part $(W_F)_{\mn\rho\si}$ is more conveniently handled within the Newman-Penrose formalism,
so we defer its discussion to later.
To derive the dispersion relation, we begin with the $R_\xi$ gauge by inserting a $-\frac{1}{2\xi}(\nabla\cdot A)^2$
term into the Lagrangian (\ref{Maxwell-I}).
The $R_\xi$ gauge is more general, and helps distinguishing the gauge mode from the physical modes.
By setting $g_{\mn}=\eta_{\mn}$ and performing partial integration, we then extract the wave equation,
\bea\label{CPT-E-WaveJ}
\left[\Box\eta^{\nu\si}-(1-\frac{1}{\xi})\prt^{\nu\si}+2(\bar{k}_F)^{\mn\rho\si}\prt_\mu\prt_{\rho}\right]A_\si=j^\nu,
\eea
from which we obtain the solution
\bea
A^\si(x)=\int_{C}\frac{d^4p}{(2\pi)^4}G_F^{\si\nu}\tilde{j}_{\nu}(p)e^{i p\cdot x},
\eea
where $\tilde{j}_{\nu}(p)$ is the Fourier transformation of $j_{\nu}(x)$ in position space and $G_F^{\si\nu}(p)=-[S(p)^{-1}]^{\si\nu}$ is the Green function of Eq. (\ref{CPT-E-WaveJ}).
The matrix $S(p)$ is defined as
\bea&&
\hspace{-5mm}S^{\mn}(p)\equiv p^2\eta^{\mn}-(1-\frac{1}{\xi})p^\mu p^\nu+2(\bar{k}_F)^{\mu\rho\nu\si}p_\rho p_\si.
\eea
By the method in Ref \cite{CLP2012}, the inverse matrix of $S(p)$ can be obtained formally as
\bea&&
\hspace{-5mm}G_F=\frac{1}{\mathrm{det(S)}}\left[(\frac{1}{3}[S^3]+\frac{1}{6}[S]^3-\hf[S^2][S])\mathbbold{1}
-S^3\right.\nn&&\left.
+[S]S^2-\hf([S]^2-[S^2])S\right],
\eea
where $[S^n]=\mathrm{tr}(S^n)$ and $\mathbbold{1}$ is the identity $4\times4$ matrix.
The dispersion relation can be obtained from the vanishing of the denominator of $G_F$,
\bea\label{detSflat}
\mathrm{det(S)}=\frac{p^2}{\xi}Q(p)=0,
\eea
where the factor $p^2=0$ multiplied with $1/\xi$ is the gauge mode, which can be confirmed by replacing
the Fourier transformation of $A^\nu(x)$,
$\tilde{A}^\nu(p)$, by $\tilde{A}^\nu(p)+p^\nu$, and the gauge invariance imposes that
$S^{\mu}_{~\nu}(p)p^\nu=p^2/\xi=0$.
The two physical modes and one longitudinal mode is given by the right factor multiplying $\frac{p^2}{\xi}$
in Eq. (\ref{detSflat}),
\bea&&\label{PhyModes}
Q(p)=(p^2)^3+2\{[K](p^2)+([K]^2-[K^2])\}p^2\nn&&
~~~~~~~+f(K)=(p^2)^2\{p^2+2[K]\}=0,
\eea
where $[K]\equiv[\Sigma_{i=1}^3(k_F)^{i\mu i\nu}-(k_F)^{0\mu0\nu}]p_\mu p_\nu$
and $f(K)=\frac{4}{3}[[K]^3-3[K][K^2]+2[K^3]]$.
Note that we have assumed a linear approximation of the LV coefficient $k_F$ in the second line of
Eq. (\ref{PhyModes}), which can be obtained either by direct calculation of $\mathrm{det}[S]$
or by using the formula provided in Ref. \cite{CLP2012}.
Interestingly, Eq. (\ref{PhyModes}) remains valid when replacing $\eta_{\mn}$ by $g_{\mn}$,
for detailed calculations, see Appendix \ref{detDRkF}.
Imposing current conservation on Eq. (\ref{CPT-E-WaveJ}) in the momentum space gives $\frac{p^2}{\xi}p\cdot\tilde{A}(p)=0$,
which shows that the physical modes are transversal, satisfying $p\cdot\tilde{A}=0$.
In the linear approximation of the $k_F$ term,
one solution with $p^2=0$ corresponds to the longitudinal mode, while the other two,
$p^2=0$ and $p^2+2[K]=0$, correspond to the two transversal modes, respectively.
It is also important to note that for CPT-even Maxwell theory, there is a symmetry $p^\mu\rightarrow-p^\mu$,
which allows us to identify $-\omega(-\vec{p})$ as the negative energy solution for each mode.

The stress-energy tensor for the CPT-even theory is
\bea\label{EgaEMT}&&
\hspace{-8mm}\theta_E^{\mn}=\theta_0^{\mn}+(k_F)^{\mu\rho\al\be}F^{\nu}_{~\rho}F_{\al\be}
-\frac{\eta^{\mn}}{4}F^{\al\be}(k_F)_{\al\be\rho\si}F^{\rho\si}.
\eea
If adding current interaction, there will be additional terms
$\eta^{\mn}(j\cdot A)+\chi^{\kappa\mu\al\be}A^\nu\prt_\kappa F_{\al\be}$ present in $\theta^{\mn}$,
and the 4-derivatives of the stress-energy tensor gives
\bea\label{EMTConsV}
\prt_\mu\theta^{\mn}=A_\mu\prt^\nu j^\mu,
\eea
which represents the conservation law in the absence of the source term $j^\mu A_\mu$.
Unlike the CPT-odd stress-energy tensor in Eq. (\ref{OgaEMT}),
the CPT-even term in Eq. (\ref{EgaEMT}) is manifestly gauge invariant if the source term is absent.
An interesting observation is that, even after applying the Belinfante symmetrization procedure,
both $\theta_E^{\mn}$ and $\theta_O^{\mn}$ are no longer symmetric.
This is a general feature of LV theory \cite{SMEg},
where the absence of angular momentum conservation prevents the stress-energy tensor from being symmetrized.
In fact, Eq. (\ref{EMTConsV}) can be viewed as a specific instance of the generic case presented
in Eq. (10) in Ref. \cite{SMEg},
though here $J^\mu=j^\mu$ represents the conserved current, not necessarily tied to
the LV background fields.

\subsection{The Maxwell equations in curved spacetime}\label{curveMaxw}
To make life simpler, we assume that the background geometry is nearly unaltered
and takes the form of pseudo-Riemann geometry, while the test particle, such as photon, experiences LV corrections.
Although this assumption may be inconsistent when matter back-reactions are considered,
we can treat the genuine LV gravitational effects as higher-order terms compared to the LV matter effects.
Thus the background geometry is still governed by the Einstein equations.

Furthermore, we assume the Lorentz symmetry is spontaneously broken and ignore the fluctuations of background fields,
as they have been subjected to stringent experimental constraints \cite{dataTable}
and may be regarded as higher-order effects.
The background observer tensor fields can be decomposed as
\bseq
\bea&&
({k}_{AF})^{\mu}=(\bar{k}_{AF})^{\mu}+(\tilde{k}_{AF})^{\mu},\\&& (k_F)^{\mn\rho\sigma}=(\bar{k}_F)^{\mu\nu\rho\sigma}+(\tilde{k}_F)^{\mu\nu\rho\sigma},
\eea
\eseq
where $(\tilde{k}_F)^{\mu\nu\rho\sigma}$ and $(\tilde{k}_{AF})^\mu$ are fluctuations and will be ignored,
and the background vacuum expectation values (vev) $(\bar{k}_F)^{\mu\nu\rho\sigma}$, $(\bar{k}_{AF})^{\mu}$
are assumed to be effectively spacetime-independent.
Note we cannot treat background vev fields as constant fields in a genuinely curved spacetime \cite{SMEg}.
However, we may suppose these vev fields varying only in a length scale $\lambda_\mathrm{LV}$
much larger than the photon wavelength
or any characteristic length scale of the physical system we considered.
In fact, the cosmological observations from Planck and WMAP on the polarization plane rotating angle $\beta={0.342^\circ}^{+0.094^\circ}_{-0.091^\circ}$ \cite{CosBire2022} yield very stringent constraints on $|k_{AF}|\leq10^{-44}$GeV at $95\%$ CL \cite{CMBLV-N}, which gives a natural length scale $\lambda_\mathrm{LV}\simeq\frac{1}{|k_{AF}|}\ge6.39\times10^5$Mpc.
Therefore, we can neglect the derivatives of $(\bar{k}_F)^{\mu\nu\rho\sigma}$ and $(\bar{k}_{AF})^{\mu}$
in the following analysis,
as the relevant length scale is significantly smaller than $\lambda_\mathrm{LV}$.
In the Lorenz gauge $\nabla\cdot A=0$, which corresponds to setting $\xi=0$, the inhomogeneous Maxwell equations (\ref{Maxwell-LVG}) reduce to
\bea\label{LVEMeom1}&&
\hspace{-3.6mm}\Box_gA^\nu-R_\la^{~\nu}A^\lambda+2(\bar{k}_F)^{\mn\rho\si}\nabla_\mu\nabla_{\rho}A_\si
\nn&&
+2\epsilon^{\mu\nu\rho\si}(\bar{k}_{AF})_\mu\prt_\rho A_\si
=-j_e^\nu.
\eea
Assuming the short-wavelength approximation (SWA) \cite{Chat22-MTW}(also known as the optical approximation),
where the characteristic electromagnetic wavelength $\lambda_c$
is much smaller than the minimal of the typical spacetime curvature radius $\mathcal{R}^{-\frac{1}{2}}$
and the typical length scale of the wave front $\lambda_0$, \ie,
$\lambda_c\ll\mathrm{min}(\mathcal{R}^{-\frac{1}{2}},\lambda_0)$,
we may neglect the curvature term $-R_\la^{~\nu}A^\lambda$ in the subsequent calculations.
Moreover, we decompose the vector potential $A^\mu$ into amplitude and phase factors
\bea
A^\mu=\mathcal{A}^\mu\exp[i\frac{S}{\epsilon}],\nonumber
\eea
where $\epsilon$ serves as a bookkeeping parameter to keep track of the order of expansions,
playing a role similar to $\hbar$ in the WKB approximation.
Note here the zero-th order term is $\mathcal{O}(\epsilon^{-2})$ and the third-order term
corresponds to $\mathcal{O}(\epsilon^0)$.
Formally, we retain the curvature term and up to the third-order expansion
in $\epsilon$, \ie, $\mathcal{O}(\epsilon^0)$. Then the equation (20) reduces to
\begin{widetext}
\bea&&
-\frac{1}{\epsilon^2}\{\mathcal{A}^\nu{p^2}+2(\bar{k}_F)^{\mn\rho\sigma}\mathcal{A}_\si{p_\mu p_\rho}
-2i\epsilon[(\mathcal{A}^\nu_{~;\rho}p^\rho+\frac{p^\rho_{~;\rho}}{2}\mathcal{A}^\nu)
+2(\bar{k}_F)^{\mn\rho\sigma}(\mathcal{A}_{\si;(\rho}p_{\mu)}+\frac{p_{\mu;\rho}}{2}\mathcal{A}_\si)
+\epsilon^{\mn\rho\si}(\bar{k}_{AF})_\mu\mathcal{A}_{\si}p_\rho]\nn&&
-\epsilon^2[\Box_g\mathcal{A}^\nu-R^\nu_{~\lambda}\mathcal{A}^\lambda+2(\bar{k}_F)^{\mn\rho\sigma}
\mathcal{A}_{\si;\mu;\rho}+2\epsilon^{\mn\rho\si}(\bar{k}_{AF})_\mu\mathcal{A}_{\si;\rho}]\}=-j_e^\nu,
\eea
where $p^\mu=g^{\mn}S_{;\nu}$ is normal (and tangent if $p^2=0$) to the equiphase surface $S(x)=C.$,
which represents the wave-front. If further expanding the amplitude as $\mathcal{A}^\nu=\sum_{n=0}\epsilon^n\bbalpha_n^\nu$, we can get
\bea\label{orderBoMaxw}&&
\mathcal{O}(\frac{1}{\epsilon^2}):
\bbalpha_0^\nu{p^2}+2(\bar{k}_F)^{\mn\rho\sigma}(\bbalpha_0)_\si{p_\mu p_\rho}=0,\nn&&
\mathcal{O}(\frac{1}{\epsilon}):
[(\bbalpha_0)^\nu_{~;\rho}p^\rho+\frac{p^\rho_{~;\rho}}{2}\bbalpha_0^\nu]
+2(\bar{k}_F)^{\mn\rho\sigma}[(\bbalpha_0)_{\si;(\rho}p_{\mu)}+\frac{p_{\mu;\rho}}{2}(\bbalpha_0)_\si]
+\epsilon^{\mn\rho\si}(\bar{k}_{AF})_\mu(\bbalpha_0)_{\si}p_\rho\nn&&
+\frac{i}{2}[\bbalpha_1^\nu{p^2}+2(\bar{k}_F)^{\mn\rho\sigma}(\bbalpha_1)_\si{p_\mu p_\rho}]=0,\nn&&
\qquad\qquad\qquad\qquad...\qquad\qquad\qquad\qquad...\qquad\qquad\qquad\qquad...\nn&&
\mathcal{O}({\epsilon}^n):\left[\Box_g\bbalpha_n^\nu{p^2}-R^\nu_{~\rho}\bbalpha_n^\rho
+2(\bar{k}_F)^{\mn\rho\sigma}(\bbalpha_n)_\si{p_\mu p_\rho}+2\epsilon^{\mn\rho\si}(\bar{k}_{AF})_\mu(\bbalpha_n)_{\si}p_\rho\right]
+2i\{[(\bbalpha_{n+1})^\nu_{~;\rho}p^\rho+\frac{p^\rho_{~;\rho}}{2}\bbalpha_{n+1}^\nu]\nn&&
+2(\bar{k}_F)^{\mn\rho\sigma}[(\bbalpha_{n+1})_{\si;(\rho}p_{\mu)}+\frac{p_{\mu;\rho}}{2}(\bbalpha_{n+1})_\si]
+\epsilon^{\mn\rho\si}(\bar{k}_{AF})_\mu(\bbalpha_{n+1})_{\si}p_\rho+\frac{i}{2}[\bbalpha_{n+2}^\nu{p^2}
+2(\bar{k}_F)^{\mn\rho\sigma}(\bbalpha_{n+2})_\si{p_\mu p_\rho}]\}\nn&&
=0,\qquad\qquad\mathrm{where}~~n\ge0.
\eea
\end{widetext}
Clearly, from the leading-order equation
\bea\label{leadDR}
M(p)^{\nu\si}(\bbalpha_0)_\si=[g^{\nu\si}{p^2}+2(\bar{k}_F)^{\mn\rho\sigma}{p_\mu p_\rho}](\bbalpha_0)_\si=0,
\eea
we get the dispersion relation
$\det[M(p)^{\nu\si}]=0$,
which does not receive any correction from the $(k_{AF})^\mu$ term, and is consistent with the analysis of the axion-like electrodynamics \cite{Itin2007}.
Similarly, as in the cases of flat spacetime, $M(p)^{\nu\si}p_\si=(p^2)p^\nu$ and $M(p)^{\nu\si}p_\nu=(p^2)p^\si$,
which would vanish if we impose the gauge fixing condition $\nabla\cdot A=0$ and then $M(p)^{\nu\si}$ changes to $M(p)^{\nu\si}-p^\nu{p^\si}$.
These results arise from the gauge invariance and current conservation, respectively \cite{SMEb}.
The current conservation condition $p_\nu j^\nu=0$ implies $p_\nu A^\nu=0$, which means $A^\nu$ is orthogonal to $p^\mu$. Consequently, $\bbalpha_0^\nu$ still lies within the tangent space of the wave front, as in conventional electrodynamics.
The determinant is computed at second order of LV coefficients $k_F$, as shown in the appendix \ref{detDRkF}.
In general, it leads to two transversal physical modes with distinct dispersion relations,
resulting in vacuum birefringence \cite{Marco2015,Cosbire2011},
which contrasts with the conclusions at linear order \cite{SMEb}.

Substituting Eq. (\ref{leadDR}) into the second equation in Eqs. (\ref{orderBoMaxw}), we get
\bea\label{Nextlead}&&
\hspace{-3.6mm}[(\bbalpha_0)^\nu_{~;\rho}p^\rho+\frac{p^\rho_{~;\rho}}{2}\bbalpha_0^\nu]
+2(\bar{k}_F)^{\mn\rho\sigma}[(\bbalpha_0)_{\si;(\rho}p_{\mu)}+\frac{p_{\mu;\rho}}{2}(\bbalpha_0)_\si]
\nn&&
+\epsilon^{\mn\rho\si}(\bar{k}_{AF})_\mu(\bbalpha_0)_{\si}p_\rho=0,
\eea
where the last term involving $\bbalpha_1$ must be satisfied separately and vanishes due to the leading-order equation (\ref{leadDR}).
Since the matrix $M(p)^{\nu\si}$ is singular, the equation is automatically satisfied.
Now we decompose $\bbalpha_0^\nu=\bbalpha\,\hat{e}^\nu$ as the product of the real amplitude $\bbalpha$ and the complex unit polarization vector $\hat{e}^\nu$.
Substituting the decomposition into Eq. (\ref{Nextlead}), we can get
\bea&&\label{polarPT-a}
\hspace{-3.6mm}\hat{e}^\nu[\nabla_p\ln\bbalpha+\frac{p^\mu_{;\mu}}{2}]+[2(k_F)^{\mn\rho\si}p_{(\mu}\nabla_{\rho)}+\epsilon^{\mn\rho\si}(\bar{k}_{AF})_\mu p_\rho]\hat{e}_\si\nn&&
+\nabla_p\hat{e}^\nu+2(k_F)^{\mn\rho\si}[p_{(\mu}\nabla_{\rho)}\ln\bbalpha+\frac{p_{\mu;\rho}}{2}]\hat{e}_\si=0,
\eea
where $\nabla_p\equiv p^\rho\nabla_\rho$.
Utilizing $\bbalpha^2=(\bbalpha_0)^\nu(\bar{\bbalpha}_0)_{\nu}$, Eq. (\ref{Nextlead}) and its complex conjugate, we can get
\bea&&\label{amplitud-a}
\hspace{-12mm}\nabla_\rho[\bbalpha^2 p^\rho]=2(k_F)^{\nu\mu\rho\si}p_{\mu;\rho}(\bbalpha_0)_{(\si}(\bar{\bbalpha}_0)_{\nu)}\nn&&
+2(k_F)^{\nu\mu\rho\si}[(\bbalpha_0)_\nu\,p_{(\mu}\nabla_{\rho)}(\bar{\bbalpha}_0)_\si+c.c.].
\eea
This equation implies that unlike the LI case, the presence of the $k_F$ term makes the vector
$\bbalpha^2 p^\rho$ non-conserved.
In other words, {\it the presence of LV background fields makes the photon number no longer a conserved quantity along the
propagation direction $p^\mu$}. However, this may not be true, because in the presence of LV, a free photon
(``free” means that a particle propagates only under the influence of gravity and background fields)
does not even travel along geodesics, in other words, both $\nabla_p p^\mu\neq0$ and $\nabla_p\hat{e}^\nu\neq0$.
This can be checked by noting that, to the linear order of the $k_F$ correction, the dispersion relations read
\bea
p^2+2(k_F)^{\mu\rho~\si}_{~~\mu}p_\rho p_\si=0,\quad
p^2=0.
\eea
The seemly LI dispersion relation arises because we have adopted the linear approximation.
If higher-order LV corrections are included [see Eq. (\ref{2ndDisRkF})], both two modes exhibit LV behaviors.
The LV dispersion relations imply that $p_\mu=\prt_\mu S$ is not null, and $\nabla_p p^\mu\neq0$,
as shown by simple algebra.
So generally speaking, $p^\nu$ is not a parallel propagated null vector.
However, at least for the leading-order birefringence-free case, where $W_F^{\mn\rho\si}=0$ in $k_F$,
$p^\mu$ is still a null vector with respect to an effective metric $\tilde{g}_\mn=g_{\mn}+4(c_F)_{\mn}$ \cite{LVBHTher}.
An interesting question is that when considered in a more general setting, such as the Finsler geometry
as opposed to pesudo-Reimann geometry \cite{Marco2015,FinslerLV},
whether a more appropriate definition of the null vector for massless particles can be established \cite{LightConeFinsler}, which may allow massless particles to travel along geodesics defined in
a geometry that accommodates directional dependence.

Even one can define a null vector in the leading order of $k_F$ correction with an effective LV modified metric
and thus $\nabla_p p^\mu=0$,
the polarization vector is still not parallel-transported along itself, at least when the $k_{AF}$ term is present.
From Eqs. (\ref{polarPT-a}) and (\ref{amplitud-a}), we can get
\bea\label{polarPT-b}&&
\hspace{-4.5mm}\nabla_P\hat{e}^\nu=\epsilon^{\nu\mu\rho\si}(\bar{k}_{AF})_\mu p_\rho\hat{e}_\si+
2[p_{(\mu}\nabla_{\rho)}\ln\bbalpha+\frac{p_{\mu;\rho}}{2}]\nn&&
\cdot[(k_F)^{\nu\mu\rho\si}\hat{e}_\si-\hat{e}^\nu(k_F)^{\ga\mu\rho\si}\hat{\bar{e}}_{(\ga}\hat{e}_{\si)}]
+2(k_F)^{\nu\mu\rho\si}p_{(\mu}\nabla_{\rho)}\hat{e}_\si\nn&&
-\hat{e}^\nu(k_F)^{\ga\mu\rho\si}[\hat{\bar{e}}_{\ga}p_{(\mu}\nabla_{\rho)}\hat{e}_\si
+\hat{e}_{\ga}p_{(\mu}\nabla_{\rho)}\hat{\bar{e}}_\si],
\eea
where $\hat{\bar{e}}^\nu$ is the complex conjugate of $\hat{e}^\nu$.
Unlike the amplitude Eq. (\ref{amplitud-a}), where the CPT-odd $k_{AF}$ term plays no role
and thus consistent with the absence of a $k_{AF}$-correction to the leading-order optical approximation of Eq. (\ref{leadDR})
and the consequent dispersion relation,
the $k_{AF}$ term does contribute to the correction to the parallel displacement of the polarization vector $\hat{e}^\nu$ in Eq. (\ref{polarPT-b}).
In other words, the presence of $k_{AF}$ term definitely alters the polarization vectors during light propagation
and thus has been tightly constrained by experiments, especially the astrophysical and cosmological observations \cite{kAFCMB}.
The astrophysical constraints on the birefringent part of the $k_F$ coefficients, namely,
the $\tilde{\ka}_{e+}, ~\tilde{\ka}_{o-}$ terms, are also detailed in \cite{kFAstro}.

From the above analyses, we conclude that if only $(k_{AF})^\mu\neq0$,
we can confidently assert that, the $p_\mu$ is a null vector in the leading-order optical approximation (or SWA).
This motivates our forthcoming treatment of the Maxwell equations with Newman-Penrose formalism in the next section.
When $(k_{AF})^\mu\neq0$, the polarization vector evolves in a distinctly different manner, which may be regarded as a second order effect within the SWA.

\section{Null formalism and Maxwell equations}\label{NullTetrad}
In the absence of the $k_F$ term, $p_\mu=\nabla_\mu S$ remains a null vector at leading order of the SWA.
This motivates the treatment of the Maxwell equations using the Newman-Penrose (NP) formalism \cite{NP1962}\cite{Chand-MToBH}.
This formalism is particularly well-suited for analyzing the behavior of massless particles,
such as their asymptotic behavior or properties at large distances.
The idea underlying the NP formalism is quite simple: we might view it as a special type of tetrad or vierbein theory,
where $e_a^{~\mu}$ is associated with an idealized comoving observer moving at the speed of light,
such that its four-velocity $u^\mu=e_0^{~\mu}$ satisfies $u^2=0$.
Thus, this approach can also be regarded as a concrete realization of Einstein's thought experiment of chasing a light beam.

For a congruence of null geodesics, one may identify the tangent vector to the null geodesics as $l^\mu$,
satisfying $\nabla_l\,l^\mu=0$.
There are two opposite directions --- outgoing and ingoing --- related by time-reversal symmetry for a congruence of null geodesics:
the outgoing/future-pointed direction has already been denoted as $l^\nu$, and the ingoing/past-oriented one can be denote as $n^\nu$.
As for deviation vectors within a congruence of null geodesics, which are orthogonal to both $l^\nu$ and $n^\nu$, and span the two-dimensional transversal space, are denoted by $\xi^\nu$ and $\zeta^\nu$.
These vectors can be combined to form two complex conjugate null vectors: $m^\nu\equiv\frac{1}{\sqrt{2}}[\xi^\nu-i\zeta^\nu]$ and $\bar{m}^\nu\equiv\overline{m^\nu}$ \cite{NU1962Asym}.
Together, these four null vectors constitute a complete basis at any given point along the null curves.
We may collectively denote them as $\{\hat{E}_a=E_a^{~\mu}\prt_\mu,a=1,..,4\}=\{-\hat{n},-\hat{l},\hat{\bar{m}},\hat{m}\}$.
The notation arises naturally since we typically begin with the null 1-forms for a given metric, see Appendix \ref{Null&DFs} for further details.

The null vectors satisfy
\bea
\eta_{ab}=g_{\mn}E_a^{~\nu}E_b^{~\mu},
\eea
where $\{E_a^{~\nu},a=1,...,4\}=(-n^\nu,-l^\nu,\bar{m}^\nu,m^\nu)$ and
\bea
\eta_{ab}=\left(
            \begin{array}{cccc}
              0 & -1 & 0 & 0 \\
              -1 & 0 & 0 & 0 \\
              0 & 0 & 0 & 1 \\
              0 & 0 & 1 & 0 \\
            \end{array}
          \right)=\eta^{ab}.\nonumber
\eea
can be interpreted as the metric in tangent space, analogous to its role in standard tetrad formalism,
where $g_{\mn}e_{a}^{~\nu}e_{b}^{~\mu}=\eta_{ab}$.
The key distinction is that, in general $\mathrm{diag}[\eta_{ab}]=(-1,+1,+1,+1)$.
We stress that Latin indices starting from the beginning of the alphabet (\eg, $a,b,c,...$) range over $1,2,3,4$,
and correspond to tangent space or tetrad labeling. In contrast, Greek indices ($\mu,\nu,...$) range over $0,1,2,3$
and refer to spacetime indices.
Additionally,
Latin indices from the middle of the alphabet ($i,j,k,...$) are used for pure spatial indices ranging over $1,2,3$.
The completeness relation of the null tetrad reads
\bea
g_{\mn}=\eta_{ab}E^a_{~\mu}E^b_{~\nu}=2[m_{(\mu}\bar{m}_{\nu)}-l_{(\mu}n_{\nu)}],
\eea
where $\eta^{bc}\eta_{ab}=\delta^c_a$ and $E^a_{~\mu}=\eta^{ab}g_{\mn}E_b^{~\nu}$.
Note different references use varying notations or conventions (see Refs. \cite{NP1962,ESoEE}).
We follow the conventions of Refs. \cite{NP1962,Chand-MToBH} but with $+2$ signature instead of $-2$,
leading to some sign differences in our results.

For a source confined to a world tube with sufficient compact spatial region,
the electromagnetic fields in the wave zone satisfy
the vacuum Maxwell equations, so we may disregard the source $j^\nu$ term in Eq. (\ref{Maxwell-LVG}).
In this section, we mainly focus on the CPT-odd $k_{AF}$ term and slightly talk about the $k_F$ term.
The Faraday tensor can be written in terms of the three complex NP scalars
\bea\label{NPS-Faraday}&&
\phi_0=F_{31}=F_{\mn}m^\mu l^\nu, \phi_2=F_{24}=F_{\mn}n^\mu\bar{m}^\nu, \nn&&
\phi_1=\hf[F_{21}+F_{34}]=\hf F_{\mn}[n^\mu l^\nu +m^\mu\bar{m}^\nu].
\eea
This expression can be found in the appendix of \cite{NP1962} or in the famous textbook of Chandrasekhar \cite{Chand-MToBH}.
As mentioned above, our signature is $+2$,
so Eqs. (\ref{NPS-Faraday}) differ by a total $-1$ sign from the expressions in Refs. \cite{NP1962,Chand-MToBH}.
For compactness, the Faraday tensor can be written in terms of the Faraday 2-form
\bea&&\label{F-2formNull}
\hspace{-3.6mm}F=\hf F_{\mn}dx^\mu\wedge dx^\nu\nn&&
=\phi_0\mathbf{n}\wedge\bar{\mathbf{m}}+\bar{\phi}_0\mathbf{n}\wedge\mathbf{m}
-\phi_2\mathbf{l}\wedge\mathbf{m}-\bar{\phi}_2\mathbf{l}\wedge\bar{\mathbf{m}}
\nn&&~
+2(\mathbb{R}\mathrm{e}[\phi_1]\mathbf{l}\wedge\mathbf{n}
-i\,\mathbb{I}\mathrm{m}[\phi_1]\mathbf{m}\wedge\bar{\mathbf{m}}).
\eea
This also indicates a straightforward method for ``deriving” Maxwell's equations in terms of the three complex NP scalars.
The key insight is that {\it the 2-form $F=\hf F_{\mn}dx^\mu\wedge dx^\nu$ is coordinate-independent}.
Thus, we can express the Faraday 2-form in terms of the dual null 1-forms associated with $\{\hat{E}_a,\,a=1,2,3,4\}$ in spherical coordinates in Minkowski spacetime
and identify the NP scalars based on their spin-weight properties. For detailed derivations, see Appendix \ref{Null&DFs}.
Also note that there are 6 independent local Lorentz degree of freedom in choosing the null tetrad.
These Lorentz transformations fall into 3 categories \cite{ESoEE}:
two preserve either $l^\mu$ or $n^\mu$,
while the third keeps both the directions of $l^\mu$ and $n^\mu$ fixed and corresponds to a rotation in the transversal space spanned by $m^\mu$ and $\bar{m}^\mu$.
In terms of these differential forms, the CPT-odd action can be written compactly,
\bea
I_{O}=\int[-\hf F\wedge{^*F}+A\wedge F\wedge (k_{AF})],
\eea
where $k_{AF}\equiv(k_{AF})_\mu dx^\mu$ and ${^*F}=\frac{1}{4}\epsilon_{\mn}^{~~\,\al\be}F_{\al\be}dx^\mu\wedge dx^\nu$ is the Hodge dual of $F$.
In this form, gauge invariance is evident if $(k_{AF})=d\Lambda$,
where $\Lambda$ is a scalar function, such as an axion field.
In the scalar field scenario, we may extend the Lagrangian to include the scalar field dynamics:
$\Delta\mathcal{L}=-\hf\nabla\Lambda\cdot\nabla\Lambda-V[\Lambda]$.
This ensure that the scalar field gradient can be assigned a preferred value
with the prescribed potential $V[\Lambda]$.
We note that
\bea
A\wedge F\wedge (k_{AF})
=-d(\Lambda\,A\wedge F)+\Lambda\,F\wedge F,
\eea
so $A\wedge F\wedge (k_{AF})=\Lambda\,F\wedge F$ up to a boundary term, analogous to the dynamical Chern-Simons modified gravitational theory \cite{Cosvary2020}\cite{CSG2009}.
Interestingly, a similar genuine Chern-Simons electromagnetic theory, formulated using differential forms,
has also been demonstrated in planar electrodynamics in $1+2$-dimensional spacetime \cite{Marco2023}.

\subsection{LI Maxwell equations in Newman-Penrose form}\label{LIMaxwNull}
In terms of the differential forms, the LI Maxwell equations are
\bea\label{dFormMaxw}
dF=0,\quad d(^*F)=(^*J),
\eea
where $F=dA$.
The dual null 1-forms to $\hat{E}_a$ are denoted as $\{\mathbf{F}^a=F^a_{~\mu}dx^\mu,\,a=1,2,3,4\}=\{\bf{l},\bf{n},\bf{m},\bf{\bar{m}}\}$.
It is often more convenient to define null 1-forms from the outset.
For instance, Sachs \cite{Sachs1962}, Newman and Unti \cite{NU1962Asym} defined the covariant vector
(or 1-form) $l_\mu$ as the normal to the null hypersurface $u=\mathrm{const.}$, \ie,
$l_\mu\equiv\nabla_\mu\,u$.
They then identified $l^\nu$ as the tangent vector to the null geodesics lying on the hypersurface $u=\mathrm{const.}$.
The field strength 2-form can be expressed as
$F=\hf F_{ab}\,\mathbf{F}^a\wedge\mathbf{F}^b$, where $F_{ab}\equiv F_{\mn}E_{a}^{~\mu}E_{b}^{~\nu}$.
For exmaple, $\phi_1=F_{31}$, and so on.
The homogeneous Maxwell equation is
\bea&&
\hspace{-5mm}0=dF=\hf dF_{ab}\wedge\mathbf{F}^a\wedge\mathbf{F}^b+F_{ab}\,d\mathbf{F}^a\wedge\mathbf{F}^b\nn&&
\hspace{-2mm}=\hf\nabla_{c}F_{ab}\mathbf{F}^c\wedge\mathbf{F}^a\wedge\mathbf{F}^b+\eta^{de}F_{eb}\ga_{cda}
\mathbf{F}^a\wedge\mathbf{F}^c\wedge\mathbf{F}^b\nn&&
\hspace{-2mm}=\hf[\nabla_{c}F_{ab}+2F_{b}^{~d}\ga_{cda}]\mathbf{F}^a\wedge\mathbf{F}^b\wedge\mathbf{F}^c
\Leftrightarrow\nn&&
\nabla_{c}F_{ab}+2F_{b}^{~d}\ga_{cda}=0,
\eea
where $\ga_{cda}\equiv \eta_{de}\,E_a^{~\rho}\nabla_\rho F^e_{~\si}E_c^{~\si}$ is the Ricci rotation coefficient
or spin coefficient,
and $\{F^a_{~\mu},~a=1,...,4\}=\{l_\mu,n_\mu,m_\mu,\bar{m}_\mu\}$ are dual tetrad components to
$\{E_a^{~\mu},a=1,...,4\}=\{-n^\mu,-l^\mu,\bar{m}^\mu,m^\mu\}$.
Similarly, from the inhomogeneous Maxwell equation $d(^*F)=(^*J)$,
we get $\nabla_{c}(^*F)_{ab}+2(^*F)_{b}^{~d}\ga_{cda}=(^*J)_{cab}$.

These two equations may be combined together as a compact complex equation,
\bea\label{SDDEFT}&&
\hspace{-5mm}
dF_d=\hf\left\{\nabla_{c}[F_{(ab)}-i(^*F)_{ab}]+2[F_{b}^{~d}\right.
\nn&&~~~~
\left.-i(^*F_{b}^{~d}]\ga_{cda}\right\}
\mathbf{F}^a\wedge\mathbf{F}^b\wedge\mathbf{F}^c,
\eea
where the spin-coefficient $\ga_{cda}\equiv e_c^{~\mu}e_{d\mu;\nu}e_a^{~\nu}$ is related to the spin connection by
$\ga_{abc}=\omega_{\mu\,ab}e_c^{~\mu}=-\ga_{bac}$.
In the above discussions, we use $F^a_{~\mu},~E_a^{~\mu}$ separately to distinguish the component of null 1-form from that of null vector.
We may denote the null tetrad simply by $e_c^{~\mu}=E_c^{~\mu}$ and $e^a_{~\mu}=F^a_{~\mu}$.
The process can be reversed by expressing $F_{\mn}$ in terms of the three complex scalars $\phi_e,~e=0,1,2$,
\bea&&
\hspace{-6.6mm}F_{\mn}=2F^{[a}_{~\,\mu}F^{b]}_{~\nu}F_{ab}\nn&&
=-2[\phi_2l_{[\mu}m_{\nu]}+c.c.]-2[\phi_1+\bar{\phi}_1]n_{[\mu}l_{\nu]}\nn&&
~~
-2[\phi_1-\bar{\phi}_1]m_{[\mu}\bar{m}_{\nu]}-2[\phi_0\bar{m}_{[\mu}n_{\nu]}+c.c.],
\eea
which differs from that in Ref. \cite{FI1972} by an overall minus sign due to the difference in metric signature.

A more traditional approach begins with the component form of the Maxwell equations,
\bea\label{PreM-Maxwell}
F_{[\mn;\rho]}=0,\quad g^{\mn}F_{\nu\rho;\mu}=j_\rho.
\eea
By introducing the intrinsic derivatives
\bea
A_{a|b}\equiv\,e_a^{~\mu}A_{\mu;\nu}e_b^{~\nu}\Leftrightarrow\,A_{\mu;\nu}=e^a_{~\mu}A_{a|b}e^b_{~\nu}\nonumber
\eea
as defined in Ref. \cite{Chand-MToBH}, we can get
\bea&&
\hspace{-3mm}F_{[\mn;\rho]}=0\Rightarrow\epsilon^{\mn\rho\si}F_{\mn;\rho}=0
\Rightarrow
\,F_{[ab|c]}=0,\\&&
\hspace{-3mm}g^{\mn}F_{\mu\rho;\nu}=j_\rho\Rightarrow
\eta^{mn}F_{ma|n}=j_a\equiv\,e_a^{~\rho}j_\rho.
\eea
Next we combine all the 8 equations of $F_{[ab|c]}=0$ and $\eta^{mn}F_{ma|n}=j_a$ together
to get
\bea\label{MaxwellNP0}&&
\phi_{1|1}-\phi_{0|4}=\frac{j_1}{2},\quad
\phi_{2|3}-\phi_{1|2}=\frac{j_2}{2},\nn&&
\phi_{1|3}-\phi_{0|2}=\frac{j_3}{2},\quad
\phi_{2|1}-\phi_{1|4}=\frac{j_4}{2}.
\eea
Then with the aid of the identity
\bea&&
A_{a,b}\equiv\,e_b^{~\mu}\prt_\mu A_{a}
=\eta^{cd}\ga_{cab}A_{d}+A_{a|b},\nonumber
\eea
we can express all the intrinsic derivatives in terms of the directional derivatives and spin coefficients,
and thus get the conventional LI Maxwell equations in the NP form
\bea&&\label{LINPMaxwell}
(D+2\rho)\phi_1-(\bar{\delta}+2\al-\pi)\phi_0-\kappa\phi_2=\frac{j_1}{2},\nn&&
(\delta-2\be+\tau)\phi_2-(\Delta-2\mu)\phi_1-\nu\phi_0=\frac{j_2}{2},\nn&&
(\delta+2\tau)\phi_1-(\Delta+2\ga-\mu)\phi_0-\sigma\phi_2=\frac{j_3}{2},\nn&&
(D-2\epsilon+\rho)\phi_2-(\bar{\delta}-2\pi)\phi_1-\lambda\phi_0=\frac{j_4}{2}.
\eea
Note again all the spin coefficients such as $\rho,~\al,~\pi,~\kappa,$ \etc.
differ by an overall minus sign due to the metric signature difference,
namely, $\mathrm{diag}(\eta_\mn)=(-1,+1,+1,+1)$ and $-n^\mu l_\mu=m_\mu\bar{m}^\mu=+1$ in this work.
These spin coefficients have clear geometry significance.
For example, $\sigma$ represents the complex shear of $l^\mu$,
while $\rho$ describes the expansion of a shadow for a null congruence along $\rho$, and so on \cite{NP1962}\cite{Sachs1962}.
The $D\equiv{}l^\mu\prt_\mu,~\Delta\equiv{}n^\mu\prt_\mu,~\delta\equiv{}m^\mu\prt_\mu,
~\bar{\delta}\equiv{}\bar{m}^\mu\prt_\mu$ represent the directional derivatives along the null directions
$l^\mu,~n^\mu,~m^\mu,~\bar{m}^\mu$, respectively.

\subsection{LV Maxwell equations in Newman-Penrose form}\label{LVMaxwNull}
With the same procedure in obtaining LI Maxwell equations (\ref{LINPMaxwell}),
we can also get the NP form of the Lorentz-violating Maxwell equations.
First we transform Eq.(\ref{Maxwell-LVG}) into the null basis,
\bea\label{FrmaeMaxwell}&&
\hspace{-7mm}
\eta^{ac}F_{ab|c}+\epsilon_{abcd}(k_{AF})^aF^{cd}+(k_F)^a_{~bcd}F^{cd}_{~~|a}
=-(j_e)_b,
\eea
where we have already ignored the derivatives of the LV background tensor fields,
and $F_{ab|c}\equiv e_a^{~\mu}e_b^{~\nu}F_{\mn;\rho}e_c^{~\rho}$ is the intrinsic derivatives of Faraday tensor.
In the following, we will treat the Maxwell equations with the CPT-odd $(k_{AF})^a$ term and the CPT-even $(k_F)^a_{~bcd}$ term separately.
Before diving into tedious calculations, we'd better investigate the conformal transformation properties of these LV operators first.

The conformal property of the LI Maxwell equation is illustrated in Appendix D of the classical textbook \cite{Wald-GR}.
However, directly assigning conformal weight to $F_{\mn}$ does not seem to be a good strategy for the CPT-odd Maxwell equations, because the gauge dependent potential $A_\mu$ is involved in Eq. (\ref{Maxwell-LVG}) or the action
(\ref{Maxwell-I}), though the gauge-dependent term has been dropped in Eq. (\ref{Maxwell-LVG}) finally.
Therefore, we believe it is more reliable to assign conformal weight to $A_\mu$ from the very beginning,
otherwise there is a potential risk of inconsistency if directly assigning conformal weights to $\tilde{F}_{\mn}=\Omega^{s}F_{\mn}$ and $\tilde{A}_{\mu}=\Omega^{s_A}A_{\mu}$ separately,
\bea\label{ConfFaraday}
\tilde{F}_{\mn}=\Omega^{s_A}[F_{\mn}-2s_A\,A_{[\mu}\nabla_{\nu]}\ln\Omega].
\eea
It is evident that unless $s=s_A=0$, the inhomogeneous term $A_{[\mu}\nabla_{\nu]}\ln\Omega$ would be non-vanishing.
So we assign the conformal weight $s_A$ to $A_\mu$,
and conformal weights $f$ and $a$ to the LV coefficients $(k_F)_{\mn\rho\si}$ and $(k_{AF})_\mu$, respectively.
Please note that the conformal weight for a tensor with
lower indices is not equal to the same tensor with upper indices.
With some simple but tedious algebras, we can get
\begin{widetext}
\bea&&\label{conformalW}
g^{\mu\al}F_{\mn;\al}\rightarrow\,\tilde{g}^{\mu\al}\tilde{F}_{\mn;\al}
=\Omega^{s_A-2}g^{\mu\al}\left\{F_{\mn;\al}+(s_A+N-4)F_{\mn}\nabla_\al\ln\Omega
+2\frac{s_A}{\Omega}(\nabla_\al\nabla_{[\mu}\Omega)A_{\nu]}\right.\nn&&
~~~~~~\left.-2s_A[\nabla_\al
+(s_A+N-5)\nabla_\al\ln\Omega]A_{[\mu}\nabla_{\nu]}\ln\Omega\right\},\nn&&
g^{\mu\al}\nabla_\al[(k_F)_{\mn\rho\si}F^{\rho\si}]\rightarrow\,\tilde{g}^{\mu\al}\nabla_\al
[(\tilde{k}_F)_{\mn\rho\si}\tilde{F}^{\rho\si}]=\Omega^{s_A+f-6}g^{\mu\al}\left\{
\nabla_\al[(k_F)_{\mn\rho\si}F^{\rho\si}]+(s_A+f+N-8)\nabla_\al\ln\Omega\right.\nn&&
~~~~~~\left.\cdot(k_F)_{\mn\rho\si}[F^{\rho\si}-2s_A\,A^{[\rho}\nabla^{\si]}\ln\Omega]
-2s_A\nabla_\al[(k_F)_{\mn\rho\si}A^{[\rho}\nabla^{\si]}\ln\Omega]
\right\},\nn&&\label{ConformT}
\epsilon_{\mn\rho\si}[(k_{AF})^\mu\,F^{\rho\si}-A^\si\nabla^\mu(k_{AF})^\rho]\rightarrow
\tilde{\epsilon}_{\mn\rho\si}[(\tilde{k}_{AF})^\mu\,\tilde{F}^{\rho\si}-\tilde{A}^\si\nabla^\mu(\tilde{k}_{AF})^\rho]
=\Omega^{N+a+s_A-6}\epsilon_{\mn\rho\si}\nn&&~~~~~~
\left\{(k_{AF})^\mu[F^{\rho\si}-2s_A\,A^{[\rho}\nabla^{\si]}\ln\Omega]-A^\si[\nabla^\mu(k_{AF})^\rho+a\,(k_{AF})^\rho\nabla^\mu\ln\Omega]
\right\},
\eea
\end{widetext}
where $N$ is the dimension of spacetime and in the present context, $N=4$.
We have used the fact that $\epsilon_{\mn\rho\si}=\sqrt{-g}\bar{\epsilon}_{\mn\rho\si}\rightarrow\Omega^{N}\epsilon_{\mn\rho\si}$,
where $\bar{\epsilon}_{\mn\rho\si}$ is the totally antisymmetric Levi-Civita symbol.
It is clear that if $f=N=4$ and $s_A=0$, the first
two terms transform as $\Omega^{-2}g^{\mu\al}\{F_{\mn;\al}+\nabla_\al[(k_F)_{\mn\rho\si}F^{\rho\si}]\}$,
while the last term transforms as $\Omega^{a-2}\epsilon_{\mn\rho\si}\left[(k_{AF})^\mu\,F^{\rho\si}
-A^\si\nabla^\mu(k_{AF})^\rho\right]$ plus an additional term $-a\,\Omega^{a-2}\epsilon_{\mn\rho\si}A^\si(k_{AF})^\rho\nabla^\mu\ln\Omega$.
So if we impose $a=0$, Eq. (\ref{Maxwell-LVG}) is conformal invariant in the absence of external current,
at least at the classical level.
Interestingly, the $(k_{AF})$ coefficient has mass dimension one but conformal weight zero,
while the $(k_{F})$ coefficient has mass dimension zero but conformal weight four.
This appears to contradict with the naive expectations that a dimensionless coupling
should typically have conformal weight zero.
However, this is not entirely unexpected,
as the modified Maxwell theory operates within the framework of an effective field theory,
where scale-dependent couplings and emergent conformal properties can arise naturally.

Nevertheless, we can still interpret the $k_{AF}$- or $k_F$- modified electrodynamics as a classical field theory,
treating the background fields as classical quantities while temporarily neglecting quantum corrections.
Given the distinct properties of $k_{AF}$ and $k_F$ operators under both CPT and conformal transformations,
we will analyze them separately in the following discussion.

\subsubsection{CPT-even $k_{F}$ modification}
If only $k_F\neq0$, the inhomogeneous Maxwell equation in the frame basis reads
\bea\label{CPT-E-Maxw}&&
\hspace{-5mm}\eta^{ac}F_{ab|c}+(k_F)^a_{~bcd}F^{cd}_{~~|a}+(k_F)^a_{~bcd|a}F^{cd}
=-(j_e)_b.
\eea
We may still neglect the variation of background field and simply set $(k_F)^a_{~bcd|a}=0$.
Following the same approach as the NP formalism, where the Faraday and Riemann tensors are projected onto a set of NP scalars,
we similarly decompose $(k_F)^a_{~bcd}$ into several scalars components.
Noting that $(k_F)^a_{~bcd}$ shares the same symmetries as the Riemann tensor,
we decompose it into Ricci-like and Weyl-like parts as Eq. (\ref{kFdecom}).
For simplicity, we disregard the Ricci-like part and focus on the Weyl-like part,
projecting $(W_F)_{\mn\rho\si}$ onto 5 scalars
\bea&&\label{Weyldecomp}
\Psi_0\equiv(W_F)_{2424}=(W_F)_{\mn\rho\si}l^\mu m^\nu l^\rho m^\si,\nn&&
\Psi_1\equiv(W_F)_{2142}=(W_F)_{\mn\rho\si}l^\mu n^\nu l^\rho m^\si,\nn&&
\Psi_2\equiv(W_F)_{1342}=(W_F)_{\mn\rho\si}l^\mu m^\nu \bar{m}^\rho n^\si,\nn&&
\Psi_3\equiv(W_F)_{2113}=(W_F)_{\mn\rho\si}l^\mu n^\nu \bar{m}^\rho n^\si,\nn&&
\Psi_4\equiv(W_F)_{1313}=(W_F)_{\mn\rho\si}n^\mu \bar{m}^\nu n^\rho \bar{m}^\si,
\eea
where $(W_F)_{abcd}=(W_F)_{\mn\rho\si}e_a^{~\mu}e_b^{~\nu}e_c^{~\rho}e_d^{~\si}$ and
the set of null tetrad $\{e_a^{~\mu},a=1,...,4\}=\{-n^\mu,-l^\mu,\bar{m}^\mu,m^\mu\}$.
Note that the decomposition of the Weyl-like tensor using the NP formalism is quite natural 
and has been similarly applied to the ``W-tensor", the non-metric part
of the constitutive tensor, to classify and study optical properties
as a generalization of transformation optics \cite{DGAOE17}.
Additionally, similar ideas involving antisymmetric bivectors constructed from null vectors have been employed to
test the equivalence principle \cite{EPphotonRing}.
For simplicity, we still adopt the test particle assumption (with the photon field in this context)
to avoid addressing potential consistency issues related to the symmetry breaking mechanism
\cite{SMEg} and the Einstein-Maxwell equations,
where the back reaction of the photon field $F_{\mn}$ to the spacetime metric $g_{\mn}$ would need to be considered.
Therefore we do not encounter Riemann tensor explicitly
and the use of $\Psi_a,~a=0,...,4$ will not cause any confusion here.

For short, we will not discuss any details of the derivation and only present a simple example to demonstrate that it will be interesting to
solve the $k_F$-modified Maxwell equation in the NP form.
For this purpose, we only set $\Psi_2\neq0$, as it has the following peculiarities: 1. it has spin-wight $0$,
resulting in a simple action under $\eth$ and $\bar{\eth}$; 2. it is the only scalar constructed from all the four null vectors;
3. the coupling $(k_F)^a_{~bcd}F^{cd}_{~~|a}$ exhibits special properties that make the NP equations resemble their LI counterparts,
suggesting that they may be separable with the Teukolsky approach \cite{Teukolsky1973}.
The NP equations with only $\Psi_2\neq0$ are given below
\bea&&\label{kFPsi2NPEq}
\hspace{-4.8mm}(D+2\rho\,k_E)\phi_1-(k_E\,\bar{\delta}+2\al\,k_E-\pi)\phi_0-\kappa\phi_2=\frac{j_1}{2},\nn&&
\hspace{-4.8mm}(k_E\,\delta-2k_E\,\be+\tau)\phi_2-(\Delta-2\mu\,k_E)\phi_1-\nu\phi_0=\frac{j_2}{2},\nn&&
\hspace{-4.8mm}(\delta+2\tau\,k_E)\phi_1-(k_E\,\Delta+2\ga\,k_E-\mu)\phi_0-\sigma\phi_2=\frac{j_3}{2},\nn&&
\hspace{-4.8mm}(k_E\,D-2\epsilon\,k_E+\rho)\phi_2-(\bar{\delta}-2\pi\,k_E)\phi_1-\lambda\phi_0=\frac{j_4}{2},\nn
\eea
where $k_E\equiv1+\frac{\Psi_2}{2}$.
In light of Eq. (\ref{kFPsi2NPEq}),
the presence of LV only slightly modifies the structure of NP equation by shifting the relevant differential operators of spin coefficients from $1$ to $k_E$.
In Minkowski and Schwarzschild spacetime, where the only nonzero spin coefficients are shown in Eq. (\ref{spincoeff}),
the equation can be further simplified.
Since the conformal properties of the $(k_F)_{\mn\al\be}$ tensor are identical to those of the metric product $g_{\mu[\al}g_{\be]\nu}$
[both share the same conformal weight $s=4$, as discussed below Eq. (\ref{ConformT})],
its behavior is quite different from the CPT-odd $k_{AF}$ term.
Consequently, we may expect that the asymptotic behavior of the CPT-even modified Maxwell equations remains qualitatively similar to the LI case, \ie, $\phi_a\sim\mathcal{O}(r^{-(3-a)}),~a=0,1,2$.
Nevertheless, the CPT-even case remains worth exploring, as it may still introduce some quantitative differences
compared to the LI scenario. However, this requires further investigation and is beyond our present scope.
We may leave it into future study.

\subsubsection{CPT-odd $k_{AF}$ modification}
If only the CPT-odd coefficient $k_{AF}\neq0$, the inhomogeneous Maxwell equations in the null basis reduce to
\bea\label{CPT-O-Maxw}&&
\eta^{ac}F_{ab|c}+\epsilon_{abcd}\,k^aF^{cd}=-(j_e)_b,
\eea
where for notational simplicity, we let $k^a\equiv(k_{AF})^a$. Please do not confuse it with the wave vector,
which is denoted by $p$ in this work.
For simplicity, we assume the background geometry is either flat or Schwarzschild,
both of which exhibit spherical symmetry.
Additionally, we assume that $(k_{AF})^\mu=((k_{AF})^t,(k_{AF})^r,0,0)$ in the preferred reference frames.
Any nonzero $k^3=(k_{AF})^\mu\bar{m}_\mu$ or $k^4=(k_{AF})^\mu{m}_\mu$ would introduce dependence on polar and azimuthal angle, thereby breaks the spherical symmetry.
Under these assumptions, the NP equations (temporarily including $k^3,~k^4$ for completeness) are given by
\begin{widetext}
\bea&&\label{CPTO-NP-Maxwell}
(D+2\rho)\phi_1-(\bar{\delta}+2\al-\pi)\phi_0-\kappa\phi_2=\frac{j_1}{2}+ik^3\phi_0-ik^4\bar{\phi}_0
+ik^2(\phi_1-\bar{\phi}_1),\nn&&
(\delta-2\be+\tau)\phi_2-(\Delta-2\mu)\phi_1-\nu\phi_0=\frac{j_2}{2}-ik^1(\phi_1-\bar{\phi}_1)+ik^3\bar{\phi}_2
-ik^4\phi_2,\nn&&
(\delta+2\tau)\phi_1-(\Delta+2\ga-\mu)\phi_0-\sigma\phi_2=\frac{j_3}{2}-ik^1\phi_0-ik^4(\phi_1+\bar{\phi}_1)
-ik^2\bar{\phi}_2,\nn&&
(D-2\epsilon+\rho)\phi_2-(\bar{\delta}-2\pi)\phi_1-\lambda\phi_0=\frac{j_4}{2}+ik^1\bar{\phi}_0
+ik^3(\phi_1+\bar{\phi}_1)+ik^2\phi_2.
\eea
\end{widetext}
Inspection of the above equations reveals an interchange symmetry.
To elaborate, when we interchange the null vectors
\bea\label{interSymm}
l^\mu\leftrightarrow n^\mu,\quad m^\mu\leftrightarrow \bar{m}^\mu,
\eea
it results to a corresponding interchange of the directional derivatives
\bea
D\leftrightarrow\Delta,\qquad \delta\leftrightarrow\bar{\delta},\nonumber
\eea
the spin coefficients
\bea&&
\rho\leftrightarrow-\mu,~~~\be\leftrightarrow-\al,~~~\tau\leftrightarrow-\pi,\nn&&
\nu\leftrightarrow-\kappa,~~~\epsilon\leftrightarrow-\gamma,~~~\lambda\leftrightarrow-\sigma,\nonumber
\eea
as well as the the CPTV coefficients and the external current components
\bea
k^1\leftrightarrow k^2,~~k^3\leftrightarrow k^4;\qquad
j_1\leftrightarrow j_2,~~j_3\leftrightarrow j_4.\nonumber
\eea
As a result, the first and the second equations are fully interchanged, as are the third and fourth equations,
provided we also swap $\phi_0\leftrightarrow-\phi_2$ and $\phi_1\leftrightarrow-\phi_1$.
The exact interchange symmetry in Eqs. (\ref{CPTO-NP-Maxwell}) under transformation (\ref{interSymm})
can be understood as a direct consequence of the time-reversion invariance of the CPT-odd term $k^\mu({^*F}_{\mn})A^\nu$, provided $(k^0,\vec{k})\rightarrow(k^0,-\vec{k})$ under time reversal.
In consequent is the interchange of the outgoing and ingoing coordinates $u=t-r\leftrightarrow v=t+r$,
which in turn leads to $\phi_0\leftrightarrow-\phi_2$.
However, these coupled equations remain challenging to solve,
because the presence of LV $k^a$ couplings prevents the separation of the three NP scalars into individual decoupled equations using Teukolsky approach \cite{Teukolsky1973}.
Alternatively, the $k^a$ couplings on the right hand side may be regarded as a current induced by the electromagnetic (EM) fields, see the right hand sides of Eqs. (\ref{InhomoMaxw}).
This is reminiscent of the axion electrodynamics,
where constant external background magnetic fields act a source current,
inducing oscillating electromagnetic fields \cite{AxionED2019}.
Similarly, the LV CPT-odd $k^a$ coupling plays a role akin to axion field.
In the following, we focus on the simplified equations under the assumption of a spherically symmetric $k_{AF}$.

In this work, we focus exclusively on the external source free cases, where $j^\nu=0$ and consequently $j_a=0,~a=1,2,3,4$.
The case with $j_a\neq0$ will be addressed in a separate study \cite{HXS2024}.
In flat and Schwarzschild spacetimes and in the absence of any external electromagnetic current,
the only nonzero spin coefficients are
\bea&&\label{spincoeff}
\rho=-1/r,\quad -\al=\be=\frac{1}{2\sqrt{2}r}\cot\th, \nn&&
\mu=-\frac{g(r)}{2r}, \quad \ga=s_\ga\frac{GM}{2r^2},
\eea
where $g(r)=1,~s_\ga=0$ for flat spacetime and $g(r)=1-\frac{2GM}{r},~s_\ga=1$ for Schwarzschild spacetime.
Substituting the above equations into Eqs. (\ref{CPTO-NP-Maxwell}) yields the following equations
\bseq\label{CPTO-NP-1}
\bea&&
\hspace{-3.9mm}(\prt_r+\frac{2}{r})\phi_1-\frac{1}{\sqrt{2}r}\bar{\eth}\phi_0=ik^2(\phi_1-\bar{\phi}_1),
\label{CPTO-NP-a}\\&&
\hspace{-3.9mm}(\prt_r+\frac{1}{r})\phi_2-\frac{1}{\sqrt{2}r}\bar{\eth}\phi_1=ik^1\bar{\phi}_0+ik^2\phi_2,
\label{CPTO-NP-b}\\&&
\hspace{-3.9mm}(\prt_u-\frac{\prt_r}{2}-\frac{g(r)}{r})\phi_1-\frac{1}{\sqrt{2}r}{\eth}\phi_2=ik^1(\phi_1-\bar{\phi}_1),
\label{CPTO-NP-c}\\&&
\hspace{-3.9mm}(\prt_u-\frac{\prt_r}{2}-\frac{1}{2r})\phi_0-\frac{1}{\sqrt{2}r}{\eth}\phi_1=ik^1\phi_0+ik^2\bar{\phi}_2,
\label{CPTO-NP-d}
\eea
\eseq
where the edth operators act on a spin-weighted function ${_sf}$ with spin weight $s$ as
\bea
\left\{
\begin{array}{c}
\eth \\
\bar\eth
\end{array}
\right\} \phantom{}_sf =
\sin^{\pm s}\th(\prt_\th \pm i \csc\th\,\prt_\ph)\sin^{\mp s}\th\
\phantom{}_sf .
\eea
Upon closer examination of Eqs. (\ref{CPTO-NP-1}),
we find that the CPT-odd modification largely preserves the overall structure of the NP Maxwell equations.
Specifically: 1. the last pair of equations determine the time evolution of $\phi_0$ and $\phi_1$ while no time evolution equation exists for $\phi_2$ (\ie, $\prt_u\phi_2$);
2. the first pair of equations establish a relationship between the radial derivatives of $\phi_1$ and the angular derivatives of $\phi_0$ ($\prt_r(r^2\phi_1)/r^2$ and $\bar{\eth}\phi_0$),
as well as between the radial derivatives of $\phi_2$ and the angular derivatives of $\phi_1$ ($\prt_r(r\phi_2)/{r}$ and $\bar{\eth}\phi_1$).
In other words, {\it the first pair of equations serve as constraint equations for the fields on a given hypersurface, whereas the latter pair describe the time evolution of the system away from the null hypersurface}.
This observation implies that we can still treat $\phi_2^0(u,\th,\phi)$ as the news function, allowing us to determine the time dependence of the remain NP scalars, $\phi_1,~\phi_0$.

\section{Formal solutions of CPT-odd Maxwell equations}\label{FormalSol}
Given $\phi_0(r,\th,\phi)$ on a null hypersurface $u=u_0$, the first pair of equations in Eqs. (\ref{CPTO-NP-1})
allow us to directly determine the radial dependence of the formal solutions as below
\bseq\label{InteComplex}
\bea\label{InteComplex0}&&
\hspace{-5mm}\phi_0=\phi_0(r,\th,\phi),\\&&\label{InteComplex1}
\hspace{-5mm}\phi_1=\phi_1^\mathrm{sLI}+\phi_1^\mathrm{LV},\\&&
\hspace{-5mm}
\phi_1^\mathrm{sLI}=\frac{\phi_1^0(\th,\phi)}{r^2}+\frac{1}{r^2}\int^r\frac{dr'r'}{\sqrt{2}}
\bar{\eth}\phi_0,\nn&&
\hspace{-5mm}
\phi_1^\mathrm{LV}=-2k^2\left(\frac{\mathrm{Im}[\phi_1^0]}{r}+\int^r\frac{dr'}{r^2}\int^{r'}
\frac{d\tilde{r}\tilde{r}}{\sqrt{2}}\,\mathfrak{D}\phi_0(\tilde{r})\right),
\nn&&\label{InteComplex2}
\hspace{-5mm}\phi_2=\frac{\phi_2^0(\th,\phi)}{r}e^{ik^2r}+\frac{e^{ik^2r}}{r}\int^r\,dr'e^{-ik^2r'}
\left[\frac{\bar{\eth}\phi_1(r')}{\sqrt{2}}\right.\nn&&
\left.
+ir'k^1\bar{\phi}_0(r')\right],
\eea
\eseq
where we have defined $\mathfrak{D}\phi_0\equiv[(\prt_\th+\cot\th)\mathrm{Im}[\phi_0]-\frac{\prt_\phi\mathrm{Re}[\phi_0]}{\sin\th}]$,
$\bar{\eth}\phi_0=(\prt_\th-\frac{i}{\sin\th}\prt_\phi+\cot\th)\phi_0$ and $\bar{\eth}\phi_1=(\prt_\th-\frac{i}{\sin\th}\prt_\phi)\phi_1$.
Moreover, we have suppressed the angular variables $\th,~\phi$ in the integration functions,
except for the integration radial-constant functions, such as $\phi_a^0(\th,\phi),~a=1,2$.
Here $k^1=\frac{g(r)}{2}k^t+\frac{k^r}{2}$ and $k^2=k^t-\frac{k^r}{g(r)}$.
Although the three equations above provide only formal solutions, they still offer valuable insights.

First consider $\phi_2$, which represents the original radiation mode in LI cases.
It receives LV corrections from the Chern-Simons coupling $k_a{^*F}^{ab}A_b$,
not only through the phase factor $e^{ik^2r}$,
which alters the radiation frequency as a result of the kinematic LV effect
via the modified dispersion relation $\omega=|\vec{p}|(1\pm\frac{2k_{AF}^t}{|\vec{p}|})^{1/2}$,
but also through modifications to the amplitude and the large distance behavior.
These arise from the integration of the functions
$\frac{\bar{\eth}\phi_1^\mathrm{LV}(r')}{\sqrt{2}}+ik^1\,r'\bar{\phi}_0(r')$ in the square bracket in Eq. (\ref{InteComplex2}).
Note $\phi_1^\mathrm{LV}$ contains a term which may induce a term $-\sqrt{2}k^2\frac{\bar{\eth}\mathrm{Im}[\phi_1^0]\ln{r}}{r}$ in $\phi_2$.
However, this term is forbidden, as it would lead to an infinite radiation flux, which will be shown later.
To avoid this issue, we set $\mathrm{Im}[\phi_1^0]=0$.
In fact, for radiation induced by a charged particle,
imposing $\mathrm{Im}[\phi_1^0]=0$ corresponds to eliminating the static radial magnetic field,
since $\mathrm{Im}[\phi_1^0]\propto \vec{B}^r$, see appendix \ref{Null&DFs}.
This is reasonable constraint, as {\it a nonzero $\vec{B}^r$ would imply the existence of a magnetic monopole}
at the center, a scenario that is incompatible with our theoretical framework, which permits only CPT-odd modifications.
Given our assumption of a spherically symmetric distribution of the CPT-odd coefficients,
there must be a particular preferred frame in which the EM field is spherically symmetric.
For charged particles acting as sources, the preferred frame naturally corresponds to the rest frame of the
centre of charges.

To clarify the radial dependence of the three complex scalars $\phi_a, a=0,1,2$, we expand $\phi_a$ in powers of $1/r$. If assuming the ingoing radiation takes the conventional form ${\phi}_0\equiv\sum_{n=1}\frac{{\phi}_0^{n-1}(\th,\phi)}{r^{n+2}}$,
which ensures a finite energy flux since ${\phi}^0_0\sim\mathcal{O}(r^{-3})$, by substituting ${\phi}_0$ into Eq. (\ref{InteComplex1}), we obtain
\bseq
\bea&&
\hspace{-3mm}\label{phi1-LIexp}
\phi_1^{\overline{\mathrm{sLI}}}=\frac{\phi_1^0(\th,\phi)}{r^2}
-\frac{1}{\sqrt{2}}\sum_{n=1}\frac{\bar{\eth}{\phi}_0^{n-1}}{n\,r^{n+2}},\nn&&\label{phi1-LVexp}
\hspace{-3mm}
\phi_1^{\overline{\mathrm{LV}}}=\sqrt{2}k^2\left[\frac{\mathfrak{D}\phi_0^0\ln{r}}{r^2}-\sum_{n=1}\frac{\mathfrak{D}\phi_0^n}{n(n+1)r^{n+2}}
\right],\nonumber
\eea
\eseq
where the superscript``$\overline{\mathrm{sLI}}$" in $\phi_1^{\overline{\mathrm{sLI}}}$ indicates that it may still contain LV contributions, though its $1/r$-decay behavior differ only quantitatively from its LI counterpart.
The overline above the superscripts is used to distinguish
the naive expansion of $\phi_1$ from the logarithmic expansions in Eq. (\ref{phi1-log-exp}).
Not as a surprise, $\phi_0^n$ may also contain LV contributions.
Most strikingly, $\phi_1^{\overline{\mathrm{LV}}}\supset\mathcal{O}(\ln{r}/r^2)$ contains a logarithmic term,
indicating that {\it the asymptotic behavior of $\phi_a$ necessarily deviates from the LI peeling property at infinity} \cite{Sachs1962,NP1962}.
This suggests that $\phi_a$ should be expressed using polyhomogeneous expansions:
\bea\label{phi0-log-exp}
\phi_0
=\sum_{n}\left[\frac{{\phi}_0^{n-1,0}}{r^{n+2}}+\sum_{m=1}^{n+l}\frac{\ln^m{r}}{r^{n+2}}{\phi}_0^{n-1,m}\right],
\eea
where $\ln^i{r}\equiv(\ln{r})^i$ and $l$ is a positive integer to be determined later.
Substituting Eq. (\ref{phi0-log-exp}) into Eq. (\ref{InteComplex1}) yields $\phi_1(r)$,
which, when substituted into Eq. (\ref{InteComplex2}), allows us to determine $\phi_2$ in principle.
It is important to note that for any fixed integer $n$,
the scalars $\phi_a,~a=0,1,2$ can contain an infinite number of terms due to the
logarithmic functions, which have the property
$\lim_{r\rightarrow+\infty}\frac{(\ln{r})^m}{r^n}=0$ for any positive but finite integer $m$, provided $n\ge1$.

\begin{widetext}
The simplest scenario occurs when the logarithmic series is truncated, meaning $m\le n+l$, where $l$ is a fixed constant integer.
So with Eq. (\ref{phi0-log-exp}), we have
\bseq\label{phi1-log-exp}
\bea&&\label{phi1-LI-log-exp}
\hspace{-6mm}
\phi_1^\mathrm{sLI}=\frac{\phi_1^0(\th,\phi)}{r^2}
-\frac{1}{\sqrt{2}}\sum_{n=1}\left[\frac{\bar{\eth}{\phi}_0^{n-1,0}}{n\,r^{n+2}}
+\sum_{m=1}^{n+l}\frac{\bar{\eth}{\phi}_0^{n-1,m}}{n^{m+1}\,r^{n+2}}
\sum_{j=0}^m\frac{m!}{j!}n^j\ln^j{r}\right],\\&&\label{phi1-LV-log-exp}
\hspace{-6mm}
\phi_1^\mathrm{LV}=\sqrt{2}k^2\left[\frac{\mathfrak{D}\phi_0^{0,0}\ln{r}}{r^2}
-\sum_{n=1}\frac{\mathfrak{D}\phi_0^{n,0}}{n(n+1)r^{n+2}}
+\sum_{m=1}^{1+l}\frac{\mathfrak{D}\phi_0^{0,m}}{r^{2}}\sum_{j=1}^{m+1}\frac{m!}{j!}\ln^{j}{r}\right]
\nn&&
-\sqrt{2}k^2\sum_{n=1}\frac{1}{r^{n+2}}\sum_{m=1}^{n+1+l}\frac{m!\,\mathfrak{D}\phi_0^{n,m}}{(n+1)^{m+1}}
\sum_{j=0}^m\frac{(n+1)^j}{j!}\sum_{i=0}^j
\frac{j!}{i!}\frac{n^i\ln^i{r}}{n^{j+1}},
\eea
\eseq
\end{widetext}
where $\phi_1^0(\th,\phi)\in\mathbb{R}$ and $\phi_1^\mathrm{LV}\in\mathbb{R}$, as expected.
In the derivation, we have also used an important integral
\bea
{f_\mathrm{ln}}^{n,m}(r)=\int^r{d\rho}\frac{\ln^m{\rho}}{\rho^{n+1}}
=\frac{-1}{r^n}\sum_{j=0}^m\frac{m!}{j!}\frac{n^j}{n^{m+1}}\ln^j{r}.
\eea
The first term, proportional to $\phi_1^0$, along with the summation over $\bar{\eth}{\phi}_0^{n-1,0}$
in $\phi_1^\mathrm{sLI}$, corresponds to the LI expectations for $\phi_1$.
However, the explicit LV $k^2$ couplings in $\phi_1^\mathrm{LV}$ introduce lower-order terms that
differ significantly from the leading LI term.
Specifically, while the LI contribution begins at $\mathcal{O}(\frac{1}{r^2})$,
{\it the presence of LV shifts the leading order of $\mathrm{Re}[\phi_1]$ to $\mathcal{O}(\frac{\ln{r}}{r^2})$.
}

As for $\phi_2$, the phase factor $e^{ik^2r}$ in Eq. (\ref{InteComplex2}) obstructs direct integration.
This phase factor originates from the $ik^2\phi_2$ term in Eq. (\ref{CPTO-NP-b}), which, in turn,
arises because we have employed the null tetrad adapt to the LI lightcone structure.
However, due to CPTV, the lightcone structure is modified according to the dispersion relation in Eqs. (\ref{CPTV-DRLC}).
Taking the timelike case of $k^\mu=(k^t,\vec{0})$ as an example, the phase factor can be expressed as
\bea\label{phase}
e^{i[\vec{p}\cdot\vec{r}-\omega(\vec{p})t]}
=e^{-i\,\omega(\bar{p})\,u}e^{i\,\delta\bar{p}\,r}=e^{-i\,\omega(\bar{p})\,u}e^{\pm i\,k^t\,r},
\eea
where $u\equiv t-r$, $\bar{p}\equiv|\vec{p}|$, $\delta\bar{p}\equiv\bar{p}-\omega(\bar{p})\simeq\mp k^t$
for $|k^t|\ll\bar{p}$.
Here, we have assumed that $\vec{p}\parallel\vec{r}$, which leads to the additional phase factor $e^{\pm ik^t\,r}=e^{\pm ik^2\,r}$ for $k^\mu=(k^t,\vec{0})$.
In other words, the factor $e^{ik^2r}$ arises because the LI null tetrad $\{l,n,m,\bar{m}\}$ used to formulate the NP form Maxwell equations (\ref{CPTO-NP-Maxwell}),
does not precisely adapt to the hypersurface of the LV wavefront.

This poses a significant challenge, as now {\it by a proper definition of advanced or retarded ``null times"
(\ie, $u\equiv t-r$ or $v\equiv t+r$) to get ride of the $r$-dependence from the exponential phase term
is not easily achievable}.
This observation may inspire us to define $\phi_2=\tilde{\phi}_2\,e^{ik^2\,r}$ and $\phi_0=\tilde{\phi}_0\,e^{-2ik^1\,r}$ to remove the phase terms on the right hand side of Eqs. (\ref{CPTO-NP-b},\ref{CPTO-NP-d}).
However, this attempt proves futile due to the presence of $\phi_1$ on the right hand sides,
as seen from the reformulated equations
\bseq\label{phasestrip-Radia}
\bea&&\label{phasestrip-Radia-a}
\hspace{-6mm}
\prt_r[r\tilde{\phi}_2]=ik^1r\,\bar{\tilde{\phi}}_0\,e^{i\delta k\,r}+\frac{\bar{\eth}\phi_1}{\sqrt{2}}\,e^{-ik^2r},\\&&\label{phasestrip-Radia-b}
\hspace{-6mm}
\prt_u[r\,\tilde{\phi}_0]-\hf\prt_r[r\,\tilde{\phi}_0]=ik^2r\bar{\tilde{\phi}}_2\,e^{i\delta k\,r}+\frac{\eth\phi_1}{\sqrt{2}}\,e^{i2k^1r},
\eea
\eseq
where $\delta k\equiv2k^1-k^2=[g(r)-1]k^t+k^r[1+1/g(r)]$.
For sufficiently large $r$, where $g(r)=1$ and $k^r=0$ (for the case of timelike $k^\mu$),
we obtain $e^{i\delta k\,r}=1$ and $e^{-ik^2\,r}=e^{-2ik^1\,r}$.
This means that the phase factor $e^{ik^2\,r}$ associated with the outgoing mode $\phi_2$ is precisely
the opposite of the phase factor $e^{-2ik^1\,r}$ accompanying the ingoing mode $\phi_0$,
just as expected from {\it time-reversal symmetry between $\phi_0$ and $\phi_2$}.
However, the phase factors accompanying $\phi_1$ in the two equations are exactly opposite
and cannot be eliminated by the redefinition of $\phi_1$.
Moreover, the equations reveal that the LV couplings {\it $k^2$ and $k^1$ intertwine the ingoing and the outgoing modes $\phi_0$ and $\phi_2$ together}, significantly complicating the analysis of their falloff behaviors.

\subsection{A naive attempt}\label{naiveAttempt}
As a naive attempt, we may {\it temporarily ignore the phase factors} in the integral of Eq. (\ref{InteComplex2}),
since the phase of an EM wave is largely independent of its amplitude,
which is our primary focus in analyzing falloff behaviors.
However, it turns out that this naive attempt does not yield a closed system of equations
when expanding in terms of polyhomogeneous functions.
The likely reason for this failure is the same issue mentioned earlier:
the LI null tetrad used to derive the set of NP equations (\ref{CPTO-NP-1})
does not accommodate to the LV modified light wavefront.
As a result, the radial dependence cannot be cleanly separated from the exponential phase factors.
Nevertheless, several equations from the order expansion of the dynamical equations (\ref{CPTO-NP-c},\ref{CPTO-NP-d}) do lead to meaningful results, particularly at order $\mathcal{O}(1)$ and $\mathcal{O}(1/r)$.
Furthermore, in the absence of LV, this expansion correctly reduce to the closed series of expansion equations
of LI Maxwell equations.

Firstly, suppress the phase term and substitute Eq. (\ref{phi1-LI-log-exp}) into Eq. (\ref{InteComplex2}) gives
\bea\label{sLIphi-2}&&
\hspace{-4.8mm}
\phi_2^\mathrm{sLI}=\frac{\phi_2^0}{r}-\frac{\bar{\eth}\phi_1^0}{\sqrt{2}r^2}
+\sum_{n=1}\frac{1}{2n}\left[\frac{\bar{\eth}^2\phi_0^{n-1,0}}{(n+1)r^{n+2}}-\sum_{m=1}^{n+l}\frac{\bar{\eth}^2\phi_0^{n-1,m}}{2n^{m}r^{n+2}}\right.\nn&&
\left.\cdot\sum_{j=0}^m\frac{m!}{j!}\frac{n^j}{(n+1)^{j+1}}\sum_{i=0}^j\frac{j!}{i!}(n+1)^i\ln^i{r}\right],
\eea
where, once again, we use ``sLI" to denote any implicit LV term that shares the same falloff behavior as its LI counterpart.
Apart from the last multiple summation terms in the square bracket,
all other terms remain the same form as their LI counterparts in $\phi_2$.
Consequently, the leading term in $\phi_2$ is still of order $\mathcal{O}({r}^{-1})$, see the $\phi_2^0$ term.

Substituting Eq. (\ref{phi0-log-exp}) and Eq. (\ref{phi1-LV-log-exp}) into Eq. (\ref{InteComplex2}) gives the explicitly LV terms,
and it can be divided into two parts.
\begin{widetext}
The part proportional to $k^1$ in $\phi_2$ is
\bea&&
\hspace{-6mm}
\phi_2\subset
\frac{i\,k^1}{r}\int^r dr'\,r'\bar{\phi}_0(r')=-
\sum_{n=1}\frac{i\,k^1}{r^{n+1}}\left[\frac{\bar{\phi}_0^{n-1,0}}{n}\right.
\left.+\sum_{m=1}^{n+l}\frac{\bar{\phi}_0^{n-1,m}}{n^{m+1}}\sum_{j=0}^m\frac{m!}{j!}(n\ln{r})^j\right],\nonumber
\eea
while the part proportional to $k^2$ in $\phi_2$ is
\bea&&
\hspace{-6mm}
\phi_2\subset k^2\left[\sum_{n=1}\frac{\mathfrak{D}\phi_0^{n,0}}{n(n+1)^2r^{n+2}}-\frac{\mathfrak{D}\phi_0^{0,0}(1+\ln{r})}{r^2}
-\sum_{m,j=1}^{l+1,m}\frac{\mathfrak{D}\phi_0^{1,m}m!}{j!\,r^2}\sum_{i=0}^j\frac{j!}{i!}\ln^i{r}
\right.\nn&&\left.
+\sum_{n=1}\frac{1}{r^{n+2}}\sum_{m=1}^{n+1+l}\frac{m!\,\mathfrak{D}\phi_0^{n,m}}{(n+1)^{m+1}}
\sum_{j=0}^m\frac{(n+1)^j}{j!n^{j+1}}\sum_{i=0}^j\frac{j!\,n^i}{i!(n+1)^{i+1}}
\sum_{q=0}^i\frac{i!\,(n+1)^{q}\ln^q{r}}{q!}\right].\nonumber
\eea
\end{widetext}
In short, the leading order of $\phi_2$ is $\mathcal{O}(r^{-1})$, which is consistent with the finiteness requirement of stress energy tensor below Eq. (\ref{energyF}), and the analyses of vacuum $\check{\mathrm{C}}$erenkov radiation in Ref. \cite{Ralf2004}.

To proceed further, we rearrange Eq. (\ref{CPTO-NP-c}) as below,
\bea&&\label{u-dep-Eq}
\prt_u[r^2\phi_1]-\frac{\prt_r[r^2\phi_1]}{2}+s_\ga r_{_S}\phi_1+2k^1r^2\mathrm{Im}[\phi_1]=\frac{r\,\eth\phi_2}{\sqrt{2}},\nn
\eea
where $r_S\equiv 2GM$ and $s_\ga=1$ has been defined after Eq. (\ref{spincoeff}).
As already noted, $\phi_1^\mathrm{LV}$ does not contribute to any imaginary part of $\phi_1$ since $\mathfrak{D}$ defined just below Eq. (\ref{InteComplex2}) is real.
The lowest-order term in Eq. (\ref{u-dep-Eq}) is of $\mathcal{O}(\ln^j{r})$,
\bea
\sqrt{2}k^2\prt_u\left[\mathfrak{D}\phi_0^{0,0}+\sum_{m=1}^{1+l}\mathfrak{D}\phi_0^{0,m}\sum_{j=1}^{m+1}\frac{m!}{j!}\ln^j{r}\right]=0,
\eea
since there is no other term in Eq. (\ref{u-dep-Eq}) containing terms of $\mathcal{O}(\ln^j{r})$.
This equation cannot be solved unless we set all $\mathfrak{D}\dot{\phi}_0^{0,m}=0$ including $m=0$.
Here $\dot{\Psi}\equiv\prt_u\Psi$ for any function $\Psi$.
However, we cannot set $\mathfrak{D}\dot{\phi}_0^{0,0}=0$, as Eq. (\ref{phasestrip-Radia-b}) or Eq. (\ref{CPTO-NP-d}) reduces to $\dot{\phi}_0^{0,0}=\frac{\eth\phi_1^0}{\sqrt{2}}$ when LV is absent.
Thus, the lowest-order equation appears to be inconsistent, indicating the failure of the naive attempt.

However, a more detailed investigation suggests that the naive attempt may not be entirely fruitless.
Considering the equation of $\mathcal{O}(1)$ yields
\bea\label{0th-u-Eq}
\prt_u\phi_1^0\equiv\dot{\phi}_1^0=\frac{{\eth}\phi_2^0}{\sqrt{2}},
\eea
This equation is precisely the zero-th order LI equation
and is a manifestation of the electric charge conservation.
One can easily find out that
\bea&&\label{0th-u-Eqb}\hspace{-8mm}
\frac{d}{du}\int{d\sigma^2}\phi_1^0
=\int\frac{d\sigma^2}{\sqrt{2}}(\prt_\th+\cot\th+i\csc\th\prt_\phi)\phi_2^0
=0,
\eea
where $\int{d\sigma^2}\equiv\int^{2\pi}_0d\phi\int^\pi_0d\th\sin\th$
and the integration of $\eth\phi_2^0$ vanishes because
\bea&&
\int{d\sigma^2}\eth\phi_2^0=\int^{2\pi}_0d\phi\int^\pi_0d\th[\prt_\th(\sin\th\phi_2^0)+i\prt_\phi\phi_2^0]=0.\nonumber
\eea
since any nonsingular function at $\th=0,~\pi$, such as $\phi_2^0$, must satisfy the periodic condition $\phi_2^0(\th,\phi+2\pi)=\phi_2^0(\th,\phi)$,
the same holds for $\phi_2$, as it describes photon fields with integer spin.
From Eq. (\ref{0th-u-Eqb}), we identify $\phi_1^0$ as the charge aspect, which explains why
$\phi_1$ represents the Coulomb mode—not only because of this charge aspect but also because
$\phi_1$ is proportional to $\hat{r}$, characterizing the longitudinal EM mode.
A notable peculiarity is that $\mathrm{Im}[\phi]_1^0=0$ in the special frame where
the spherical symmetry assumption for $k^\mu=(k^t,k^r,0,0)$ holds.

The next-lowest order of Eq. (\ref{CPTO-NP-d}) gives
{\small
\bea&&\label{1th-u-Eq}\hspace{-6mm}
\dot{\phi}_0^{0,0}-ik^1{\phi}_0^{0,0}=\frac{\eth\phi_1^0}{\sqrt{2}}+\frac{ik^2\eth^2}{4}[\bar{\phi}_0^{0,0}-\sum_{m=1}^{1+l}\bar{\phi}_0^{0,m}\sum_{j=0}^m\frac{m!(j+1)}{2^{j+1}}],\nn&&
\eea
}
\hspace{-1.5mm}where we only retain the linear order of LV corrections
and neglect terms proportional to $k^1k^2$ and $(k^2)^2$.
This corresponds to Eq. (\ref{u-dep-Eq}) of $\mathcal{O}(1/r)$.
In reality, the two $u$-dependent equations, Eqs. (\ref{CPTO-NP-c},\ref{CPTO-NP-d}) are not independent as long as gauge invariance is preserved,
which holds in our CPT-odd Maxwell theory.
As noted in Eq. (\ref{phasestrip-Radia-b}), the second term, $-ik^1{\phi}_0^{0,0}$, represents a phase factor arising form the residual effect of the misalignment of the LI null tetrad with respect to the LV wave front.
By disregarding LV corrections proportional to $k^2$ in Eq. (\ref{1th-u-Eq}),
we see from Eq. (\ref{0th-u-Eq}) that the time evolution of the charge aspect of $\phi_1^0$ can be determined from the time dependence of $\phi_2^0$.
Subsequently, we can deduce the time evolution of ${\phi}_0^{0,0}$ from $\phi_1^0$.

The Eq. (\ref{CPTO-NP-d}) of order of $\mathcal{O}(\ln^i{r}/r^3)$ gives
\bea&&\hspace{-6mm}
\dot{\phi}_0^{0,1}-ik^1{\phi}_0^{0,1}=k^2\eth[\mathfrak{D}\phi_0^{0,0}
+\sum_{m=1}^{1+l}{m!}(\mathfrak{D}\phi_0^{0,m}-\frac{ik^2\eth^2}{4}\frac{\bar{\phi}_0^{0,m}}{2^{j}}),\nn&&
\eea
where again we ignore higher-order LV corrections and $-ik^1{\phi}_0^{0,1}$ is a phase misalignment term.
So given the u-dependence of $\phi_0^{0,0}$, we should be able to determine the time evolution of $\dot{\phi}_0^{0,1}$
if truncating the summation series by setting $l=0$.
However, this truncation does not prevent the generation of terms of $\mathcal{O}(\ln^m{r}/r^n)$ for $m\ge2$
and sufficiently large integer $n$ from the double integral in Eq. (\ref{InteComplex2}).
In other words, the naive attempt breaks down at some point due to the generation of higher-order logarithmic terms,
even though it correctly reproduces the series of equations in the absence of LV.
As an example, the Eq. (\ref{CPTO-NP-d}) of order $\mathcal{O}(1/r^3)$ gives
{\small
\bea&&\hspace{-6mm}
\dot{\phi}_0^{1,0}-ik^1{\phi}_0^{1,0}+{\phi}_0^{0,0}-\hf{\phi}_0^{0,1}=-\frac{\eth\bar{\eth}}{2}[{\phi}_0^{0,0}+{\phi}_0^{0,1}]\nn&&
\hspace{-6mm}-\frac{k^2\eth}{2}\left(\mathfrak{D}[\phi_0^{1,0}+\frac{3}{4}\phi_0^{1,1}+\frac{7}{4}\phi_0^{1,2}]
-\frac{i\eth}{6}[\bar{\phi}_0^{1,0}-\frac{5\bar{\phi}_0^{1,1}}{12}-\frac{19}{36}\bar{\phi}_0^{1,2}]\right).\nn&&
\eea
}\hspace{-0.3mm}Again we find that $\dot{\phi}_0^{1,0}$ can be determined from the time dependence of ${\phi}_0^{0,0}$
and ${\phi}_0^{0,1}$ if $k^2=0$.
However, the presence of $k^2$ and the logarithmic term $\bar{\phi}_0^{1,2}$ obstruct the
succeeding steps, even though {\it the series of equations close once LV terms are absent}.
In other words, given the time dependence of news function $\phi_2^0$, we can determine the charge aspect $\phi_1^0$,
and from $\phi_1^0$, we can sequentially obtain $\phi_0^{0,0}$, then $\phi_0^{1,0}$, and so on up to $\phi_0^{n,0}$.

\subsection{Further analyses}
The above analyses seems to indicate that the naive attempt to ignore the phase factors
caused by LV fails to produce a closed series of order expansions as the NP equations of LI Maxwell theory.
However, it is not totally fruitless, as it still yields some interesting equations,
such as charge conservation, which remains unaffected by LV, see Eq. (\ref{0th-u-Eq}).
In general, a set of equations can always be projected onto a given tetrad,
meaning that an improper choice of a set of tetrad does not invalidate the equations themselves,
though it does complicate the analyses --- much like studying a dynamic problem in an unsuitable reference frame.
A similar situation may arise here.
Nevertheless, we can still extract useful results from the exact formal integral equations (\ref{InteComplex}).
This set of equations reveal that the phase factors do contain valuable information about the falloff behaviors.

With the polyhomogeneous expansion (\ref{phi0-log-exp}),
we find that $\phi_2$ in (\ref{InteComplex2}) must contain an integral of the form
\bea&&\label{exactEe}\hspace{-6mm}
\frac{e^{ik^2r}}{r}\int^r\,dr'\frac{e^{-i\,k^2r'}}{r'^{n+1}}
=-\frac{e^{i\,k^2\,r}}{r^{n+1}}\mathrm{Ee}[n+1,i\,k^2r]\nn&&
\hspace{-8mm}
=\frac{-1}{n}\left(\frac{1}{r^{n+1}}+\frac{i\,k^2\,e^{ik^2r}}{r^n}\mathrm{Ee}[n,i\,k^2r]\right)\nn&&
\hspace{-8mm}
=-\sum_{m=1}^M(\frac{i}{k^2})^m\frac{(n+m-1)!}{n!\,r^{n+1+m}}+\mathcal{O}(r^{-(n+M+2)}),
\eea
where $\mathrm{Ee}[n+1,i\,k^2r]\equiv\int_1^{+\infty}\frac{dt}{t^{-(n+1)}}e^{-ik^2 r\,t}$ is the exponential integral function.
Clearly, the final expansion equation suggests that the perturbative expansion in the tiny LV parameter
$k^2$ may be problematic, since $k^2$ appears in the denominators of the series expansion.
In other words, {\it the integral involving the LV phase factor $e^{-ik^2 r}$ is highly nonperturbative in the tiny parameter $k^2$}.
This is not entirely unexpected.
Consider, for instance, a simple spherical monochromatic wave,
whose Taylor expansion yields
\bea
\frac{e^{i\,p\,r}}{r}=\frac{1}{r}+\sum_{n=1}^{+\infty}\frac{(i\,p)^n\,r^{n-1}}{n!}.
\eea
This expansion is meaningful only when $p\,r\ll1$ and is dominated by the first two terms, $\frac{1}{r}$ and $i\,p$, where the former represents amplitude and the latter the phase.
For large $p\,r$, 
the expansion is meaningless, since the amplitude is always dominated by the $1/r$ behavior,
rather than by higher-power terms such as $r^n$.

The second equality in Eq. (\ref{exactEe}) implies that
the LI contributions can always be separated from the exact formula (\ref{InteComplex2}),
\bea&&\label{EqualityEe}\hspace{-3mm}
\phi_2\subset
\frac{e^{ik^2r}}{r}\int^r\,dr'e^{-ik^2r'}\frac{\bar{\eth}\phi_1(r')}{\sqrt{2}}\subset\nn&&\hspace{-6mm}
-\frac{e^{ik^2r}}{2r}\int^r\,dr'e^{-ik^2r'}\sum_{n=1}\frac{\bar{\eth}^2{\phi}_0^{n-1,0}}{n\,r'^{n+2}}
=\sum_{n=1}\frac{\bar{\eth}^2\phi_0^{n-1,0}}{2n}
\nn&&\hspace{-3mm}
\cdot\left[\frac{1}{(n+1)r^{n+2}}+\frac{i\,k^2\,e^{ik^2r}}{r^{n+1}}\mathrm{Ee}[n+1,i\,k^2r]\right].
\eea
In comparison, the first term in the last equation corresponds exactly to the third term in Eq. (\ref{sLIphi-2}) of $\phi_2^\mathrm{sLI}$.
This confirms that why our naive attempt in sec. \ref{naiveAttempt} can still reproduce the LI results, 
despite completely ignoring the LV phase factors.

\begin{figure}[htbp]
  \centering
  \hspace{-0.05in}
  {\includegraphics[width=0.45\textwidth]{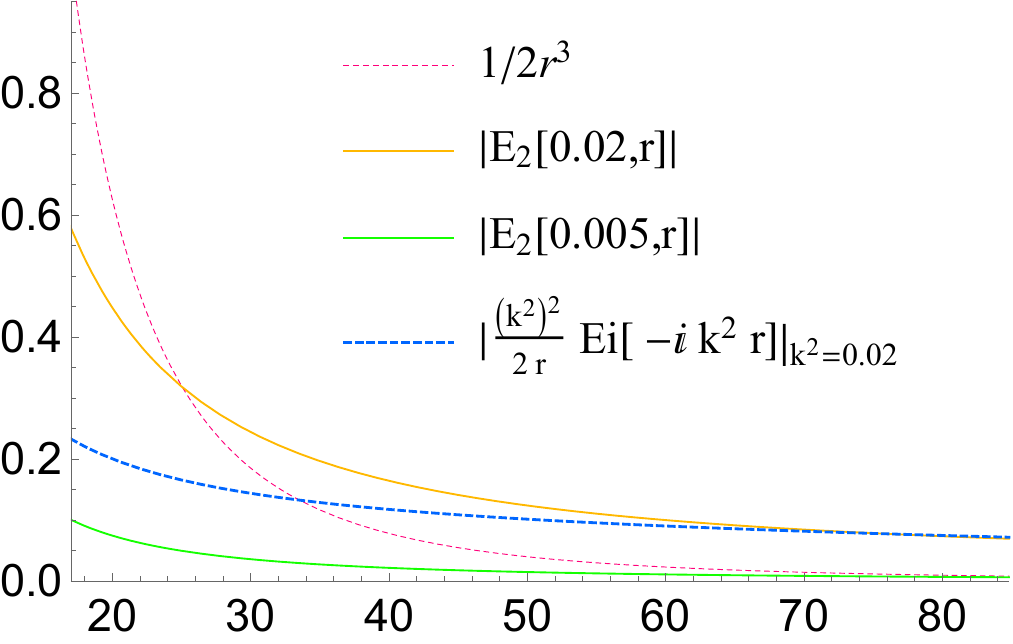}}
  \caption{Comparison of falloff behaviors.
  The red dashed curve represents the LI terms with a falloff of $\mathcal{O}(r^{-3})$,
  while the orange and green curves represent the LV corrections in Eq. (\ref{Ee3-Ei}),
  and the dashed blue curve represents the absolute value of the last LV correction,
  which is proportional to $(k^2)^2$.
  All curves are magnified by a factor $10^4$,
  and the $\mathrm{E}_2[k^2,r]$ function in the legend is defined just below Eq. (\ref{Ee3-Ei}).
  The figure clearly illustrates that the LI term dominates when $r\le\kappa/k^2$,
  where $\kappa=0.5013$ is determined as the root of the dimensionless equation of $|\mathrm{E}_2[k^2,\tilde{r}]2\tilde{r}^3|=1$ with $\tilde{r}=k^2r$.}\label{compare}
\end{figure}

Moreover, it also reveals that the LV contribution does alter the falloff behavior, though not in a straightforward way.
From the last equation in (\ref{EqualityEe}), we see the second term, proportional to $e^{ik^2r}$,
represents the LV corrections.
With the iteration relation $\mathrm{Ee}[n+1,i\,k^2r]=\frac{1}{n}(e^{-i\,k^2r}-i\,k^2r\mathrm{Ee}[n,i\,k^2r])$,
we can write any term of the form $\frac{i\,k^2\,e^{ik^2r}}{r^{n+1}}\mathrm{Ee}[n+1,i\,k^2r]$ as a sum of terms with lower power of $r^{-1}$ and higher power of $k^2$. For example,
\bea&&\label{Ee3-Ei}\hspace{-6mm}
-\frac{e^{i\,k^2r}}{r^{3}}\mathrm{Ee}[3,i\,k^2r]=\frac{-1}{2r^3}+\frac{i\,k^2}{2r^2}-\frac{(k^2)^2e^{i\,k^2r}}{2r}\mathrm{Ei}[-i\,k^2r],\nn&&
\eea
where $\mathrm{Ei}[z]\equiv-\int_z^{+\infty}\frac{e^{-t}}{t}dt$ and for later convenience,
we define the terms on the right hand side without $\frac{-1}{2r^3}$ as $\mathrm{E_2}[k^2,r]$.
It appears that we may ignore the higher power of $k^2$ since it is a small parameter.
However, in the ideal propagation case, where the photon has not been scattered away or decays \cite{photondecay2024}
in certain LV scenarios \cite{photondecay2009},
the latter higher-order term of $k^2$ can eventually dominate the former lower-order term as $r$ increases.
This behavior is clearly illustrated in Fig. \ref{compare}.
This observation has three key implications:
1. The exact integral and the iteration formula suggests that, in the asymptotic behavior at null infinity,
{\it the CPTV Maxwell theory is inherently non-perturbative, which means we cannot neglect higher-order LV corrections.}
Interestingly, $k^2=k^t-k^r$ precisely appears as the denominator in the non-perturbative polarization structure \cite{vacuoCerenRad}.
2. The seemingly lower falloff LV terms introduce a natural length scale $\lambda_k\equiv\kappa/k^2$
(where $\kappa$ is an order 1 numerical factor).
When $r\le\lambda_k$, the falloff behavior is dominated by LI results, consistent with the peeling theorem: $\phi_a^0\sim\mathcal{O}(r^{a-3}),~a=0,1,2$.
However, when $r\ge\lambda_k$, the LV corrections dominate against LI falloff behaviors,
though only for the Coulomb mode.
3. At large distances, considering only the lowest order behavior,
{\it $\phi_2$ is still governed by an $r^{-1}$ decay},
since no LV term can induce lower or even comparable falloff behaviors,
which is also expected from the physical requirement of radiation flux.
Meanwhile, {\it $\phi_1$ may exhibit LV deviations of $\mathcal{O}(k^2\ln{r}/r^2)$},
which exceed the LI behavior of $\mathcal{O}(r^{-2})$ for sufficiently large $r$.
However, since $k^2$ is strongly constrained by experiments (see table D16 in Ref. \cite{dataTable}),
$\ln{r}$ grows too slowly, and the longitudinal Coulomb mode is a near field effect,
$\phi_1$ effectively remains $\mathcal{O}(r^{-2})$. 
In short, {\it the peeling theorem remains effectively unaltered in the presence of CPTV $k_{AF}$ coupling}.

\section{Energy-momentum tensor}\label{EMTNull}
The energy-momentum tensor (EMT) in the presence of Lorentz violation is a subtle issue in curved spacetime.
One key reason is that the lack of Lorentz symmetry implies angular momentum tensor is no longer a conserved quantity.
Consequently, the traditional Belinfante procedure to obtain a symmetric EMT is obstructed
due to the absence of a conserved angular momentum current density.
For this reason, the conventional way of linking the canonical EMT with the symmetric EMT,
which is derived from the variation of the matter action with respect to the metric tensor,
is no longer valid \cite{SMEg}.
This discrepancy means that the EMT obtained from the Noether theorem in classical field theory
and the EMT from general relativity, which serves as the source of gravitational fields,
are not equivalent in the presence of non-dynamical background LV fields \cite{SMEg}.
For details, see Appendix \ref{EMTqftvsgr}.
Given this context, it is also valuable to examine the EMT of the LV-modified Maxwell theory
expressed in terms of the NP formalism.
The EMT, incorporating both CPT-odd and CPT-even corrections, is presented as below
\bea&&\label{emtA}
\Th^{\mn}_A=\left[\chi^{\mu\rho\al\be}F_{\al\be}+\epsilon^{\mu\rho\al\be}k_\al\,A_\be\right]F^\nu_{~\rho}+g^{\mn}\mathcal{L}
\nn&&
~~~~~~+A^\nu\nabla_\rho[\chi^{\rho\mu\al\be}F_{\al\be}+\epsilon^{\rho\mu\al\be}k_\al\,A_\be],
\eea
where $\chi^{\mu\rho\al\be}F^{\nu}_{~\rho}F_{\al\be}=\left[F^{\mu\rho}+(k_F)^{\mu\rho\al\be}F_{\al\be}\right]F^{\nu}_{~\rho}$.
The last term vanishes for on-shell fields outside the world tube of the external source current,
since the equation of motion gives
$\nabla_\rho[\chi^{\rho\mu\al\be}F_{\al\be}+\epsilon^{\rho\mu\al\be}k_\al\,A_\be]=j^\mu=0$.
Thus, only the first line in Eq. (\ref{emtA}) remains.
The CPT-odd or CPT-even EMT can be obtained by setting $k_F=0$ or $k_{AF}=0$, respectively.
It is evident that both CPT-odd and CPT-even EMTs remain asymmetric even after applying the Belinfante symmetrization procedure.
As previously mentioned, the absence of a conserved angular momentum current
makes it impossible to obtain a symmetric EMT through the Belinfante procedure.

In the following, we only talk about the CPT-odd EMT, $\Th^{\mn}_O=\Th^{\mn}_0+\delta{\Th}_O^{\mn}$,
where $\Th^{\mn}_0,~\delta{\Th}_O^{\mn}$ denote the LI and LV contributions, respectively.
Since $\Th_{ab}\equiv\Th_{\mn}e_a^{~\mu}e_b^{~\nu}$, we find that
\bea
(\Th_0)_{ab}\equiv2\left(
                  \begin{array}{cccc}
                   |\phi_0|^2 & |\phi_1|^2 & \phi_0\bar{\phi}_1 & \phi_1\bar{\phi}_0 \\
                   |\phi_1|^2 & |\phi_2|^2 & \phi_1\bar{\phi}_2 & \phi_2\bar{\phi}_1 \\
                   \phi_0\bar{\phi}_1 & \phi_1\bar{\phi}_2 & \phi_0\bar{\phi}_2 & |\phi_1|^2 \\
                   \phi_1\bar{\phi}_0 & \phi_2\bar{\phi}_1 & |\phi_1|^2 & \phi_2\bar{\phi}_0 \\
                  \end{array}
                \right),\nonumber
\eea
where one can easily verify that $(\Th_{ab})_0=(\Th_{ba})_0$ and that the trace vanishes, $\eta^{ab}(\Th_0)_{ab}=0$,
which reflects the conformal invariance of the classical LI Maxwell theory.
Consequently, $(\Th_0)_{ab}$ can have only 9 independent components.
The components of CPT-odd correction are listed below,
\bea&&
\hspace{-5.6mm}
(\delta{\Th}_O)_{11}=2 k^1\left(A^2\mathrm{Im}[\phi_1]-\mathrm{Im}[A^3 \phi_0]\right), \nn&&
\hspace{-5.6mm}
(\delta{\Th}_O)_{12}=2 k^2\left(A^2\mathrm{Im}[\phi_1]-\mathrm{Im}[A^3 \phi_0]\right), \nn&&
\hspace{-5.6mm}
(\delta{\Th}_O)_{13}=(\overline{\delta{\Th}}_O)_{14}=\nn&&
~~~~~=2 k^3\left(A^2\mathrm{Im}[\phi_1]-\mathrm{Im}[A^3 \phi_0]\right),\nn&&
\hspace{-5.6mm}
(\delta{\Th}_O)_{21}=2 k^1\left(A^1\mathrm{Im}[\phi_1]+\mathrm{Im}[A^4 \phi_2]\right), \nn&&
\hspace{-5.6mm}
(\delta{\Th}_O)_{22}=2 k^2\left(A^1\mathrm{Im}[\phi_1]+\mathrm{Im}[A^4 \phi_2]\right), \nn&&
\hspace{-5.6mm}
(\delta{\Th}_O)_{23}=(\overline{\delta{\Th}}_O)_{24}\nn&&
~~~~~=2 k^3\left(A^1\mathrm{Im}[\phi_1]+\mathrm{Im}[A^4 \phi_2]\right),\nn&&
\hspace{-5.6mm}
(\delta{\Th}_O)_{31}=(\overline{\delta{\Th}}_O)_{41}\nn&&
~~~~~=-i k^1\left(2A^4\mathrm{Re}[\phi_1]+A^2 \bar{\phi}_2+A^{1} \phi_0\right)\nn&&
\hspace{-5.6mm}
(\delta{\Th}_O)_{32}=(\overline{\delta{\Th}}_O)_{42}\nn&&
~~~~~=-i k^2\left(2A^4\mathrm{Re}[\phi_1]+A^2 \bar{\phi}_2+A^{1} \phi_0\right),\nn&&
\hspace{-5.6mm}
(\delta{\Th}_O)_{33}=(\overline{\delta{\Th}}_O)_{44}\nn&&
~~~~~=-i k^3\left(2A^4\mathrm{Re}[\phi_1]+A^2 \bar{\phi}_2+A^{1} \phi_0\right),\nn&&
\hspace{-5.6mm}
(\delta{\Th}_O)_{34}=(\overline{\delta{\Th}}_O)_{43}\nn&&
~~~~~=-i k^4\left(2A^4\mathrm{Re}[\phi_1]+A^2 \bar{\phi}_2+A^{1} \phi_0\right).\nonumber
\eea
A naive counting of $(\delta{\Th}_O)_{ab}$ seems indicate that the CPT-odd EMT has 14 independent components,
since $\Th^{0j}\neq\Th^{j0}$ and $\Th^{ij}\neq\Th^{ji}$ (or equivalently $\Th_{ab}\neq\Th_{ba}$).
By performing an integral over the time translation Killing vector $n^\mu$, we can obtain the energy flux
\bea&&\label{energyF}
\hspace{-3.6mm}
\mathcal{F}\propto\int{du}\,r^2d\si^2\Th_{\mn}n^\mu n^\nu=\int{du}\,r^2d\si^2[(\Th_0)_{22}+(\delta{\Th}_O)_{22}]\nn&&
=\int{du}\,r^2d\si^2\left[|\phi_2|^2+2 k^2\left(A^1\mathrm{Im}[\phi_1]+\mathrm{Im}[A^4 \phi_2]\right)\right].
\eea
In spherical coordinate, $A^1=A_r$  
and from Eq. (\ref{phi1-LI-log-exp}), we have $\mathrm{Im}[\phi_1]=\mathrm{Im}[\phi_1^\mathrm{sLI}]\sim\mathcal{O}(\ln{r}/r^3)$,
so the second term, $r^2\,A^1\mathrm{Im}[\phi_1]\sim\mathcal{O}(\ln{r}/r)$, does not contribute at null infinity ($r\rightarrow+\infty$), provided that $A^1\sim\mathcal{O}(1)$,
which is a reasonable and easily satisfied condition.
This is naturally guaranteed since the second term vanishes in the metric $ds^2=r^2(d\th^2+\sin^2\th d\phi^2)-du^2-2dudr$ and under the radiation gauge condition where $A^1=A_r=0$.
For the third term $r^2\mathrm{Im}[A^4 \phi_2]$, since $A^4=\frac{1}{\sqrt{2}r}(A_\th-\frac{i}{\sin\th}A_\phi)$ and $\phi_2\sim\mathcal{O}(1/r)$,
this term is also finite provided $A_\th,~A_\phi\sim\mathcal{O}(1)$, which is also easily satisfied.
As for the first term, which is the sole remaining term in the LI energy flux of conventional Maxwell theory,
$|\phi_2|^2\sim\mathcal{O}(r^{-2})$ automatically ensures the finiteness of $\mathcal{F}$.

The above analysis suggests that the finiteness of energy flux serves as a useful but relatively loose constraint
on the falloff behaviors of the EM fields.
It only requires that $\phi_2\sim\mathcal{O}(r^{-1})$, which is represented by the leading term, the news function $\phi_2^0$.
As in the LI case, this term is also responsible for the non-vanishing of energy flux.
However, the behavior of other terms, such as $\phi_0,~\phi_1$, still needs to be examined more closely through the corresponding Maxwell equations.
Nevertheless, from Eq. (\ref{energyF}), we find that the energy flux $\mathcal{F}$ may be slightly modified by the $k^2\mathrm{Im}[A^4 \phi_2]$ term,
since it can contribute a small but finite part at null infinity, provided $A_\th,~A_\phi\sim\mathcal{O}(1)$,
which is a natural requirement based on dimensional analysis.
In fact, the terms inside the square bracket in Eq. (\ref{energyF}) are parallel to those in $\vec{S}=\vec{E}\times\vec{B}-k^t(A^t\vec{B}-\vec{A}\times\vec{E})$,
representing the energy flux $\theta_O^{j0}$ in Eq. (\ref{OgaEMT}),
and the last two terms can be rewritten as
\bea
k^t(\vec{A}\times\vec{E}-A^t\vec{B})=k^t[\nabla\times(A^t\vec{A})-\vec{A}\times\dot{\vec{A}}].
\eea
The first term does not contribute to a large spatial integral and the second term $-k^t\vec{A}\times\dot{\vec{A}}$ contains odd time derivatives.
As a result, the integral over $k^2\mathrm{Im}[A^4 \phi_2]$, which is parallel to $\vec{A}\times\vec{E}$,
may also yield zero total radiation
when averaged over a time period, such as in the case of dipole radiation \cite{vacuoCerenRad}.

As for the CPT-even case, 
analyzing the full EMT is challenging due to the 19 independent components of $k_F$.
However, if we assume that only the Weyl-like term $W_F$ in $k_F$ is nonzero,
the energy flux will be proportional to $T_{22}=4\left[\frac{(1+\Psi_2)}{2}|\phi_2|^2-(\Psi_3\mathrm{Re}[\phi_1]-\Psi_4\phi_0)\bar{\phi}_2\right]$.
Since the leading terms of $\phi_0$ and $\phi_1$ (denoted as $\phi_0^0$ and $\phi_1^0$) generally decay more rapidly
than $\phi_2^0$, the finiteness of energy flux is automatically satisfied.
Thus, by itself it does not impose any constraint on the falloff behavior of $\phi_a$ for $a=0,1,2$.
We can only infer that the CPT-even $k_F$ correction is unlikely to introduce any qualitative deviations
in the falloff behavior at large distance compared to the conventional LI theory.

\section{Discussions}\label{Summary}
The asymptotic properties, or falloff behaviors, of massless fields at large distances
are crucial for disentangling radiation modes from the Coulomb mode.
This not only helps clarify the physical degree of freedom,
but also validates the physical significance of certain fields, such as the existence of gravitational waves.
However, in the presence of Lorentz violation (LV), these asymptotic properties have not received enough attention.
One noticeable exception is Ref. \cite{Yuri2021Asym}, where the so-called
$t$-puzzle was identified as a consequence of the non-minimal coupling $t^{\mn\al\be}W_{\mn\al\be}$ between the LV $t$ coefficient and the Weyl tensor.
This non-minimal coupling is incompatible with non-trivial static and spherically symmetric solutions other than flat spacetime.
In other words, $t$-coupling effectively constrains the asymptotic behavior of all the component of the curvature tensor including the Weyl tensor.

In this work, we review the minimal extension of the Maxwell Lagrangian within the framework of SME.
We calculate the field equations in the optical approximation for a generic curved spacetime,
and explicitly show in Eq. (\ref{polarPT-a}) that the photon’s polarization vector $\hat{e}^\nu$,
cannot be parallelly propagated along the direction of propagation, \ie, $\nabla_p\hat{e}^\nu\neq0$,
and in Eq. (\ref{amplitud-a}) that the photon flux, $\bbalpha^2 p^\rho$, is not conserved, $\nabla_\rho(\bbalpha^2 p^\rho)\neq0$.
Furthermore, we demonstrate that in the CPT-odd sector, photon's dispersion relation
and, consequently, the light cone structure remain unaltered at leading order in the optical approximation.

This partially motivates our study of the Maxwell equations with the LV corrections in the Newman-Penrose (NP) formalism.
We begin with a brief review of the NP formalism and derived the NP form of the LI Maxwell equations with both the intrinsic derivative approach \cite{Chand-MToBH}
and the coordinate-independent differential form approach.
The latter method applies to any theory expressible in differential forms, such as the CPT-odd Chern-Simons-Maxwell theory.
Next, we present the NP form of the Maxwell equations modified by the CPT-odd $k_{AF}$ and CPT-even $k_F$ coefficients separately.
Given that $k_F$ coefficients contain 19 independent degrees of freedom, for simplicity,
we focus only on one of the Weyl-like component, $\Psi_2$ [see Eq. (\ref{Weyldecomp})].
To ensure the meaningful asymptotic behavior of the LV Maxwell equations at null infinity,
we analyze the conformal transformation properties of the $k_F$ and $k_{AF}$ terms.
Since conformal invariance is crucial for preserving the causal structure determined by the metric tensor \cite{Wald-GR},
our discussion below Eq. (\ref{conformalW}) shows that, at least at the classical level,
these equations are conformal invariant provided that the $k_{AF}$ and $k_F$ coefficients can be assigned appropriate conformal weights.

We then focus mainly on the CPT-odd Chern-Simons-Maxwell theory.
For simplicity, we assume $k_{AF}^\mu=(k^t,~k^r,0,0)$ in a spherically symmetric spacetime,
such as Schwarzschild or Minkowski spacetime.
The NP form of Maxwell equations reveals that the CPT-odd couplings mix the Coulomb mode $\phi_1$, ingoing and outgoing modes $\phi_0,~\phi_2$ altogether,
even under the simple assumption of spherical symmetry, see Eq. (\ref{CPTO-NP-Maxwell}) and Eq. (\ref{CPTO-NP-1}).
This significantly complicates our analysis.
Furthermore, since projecting the CPT-odd Maxwell equations onto the LI null tetrad cannot rip off the radial coordinate from the exponential phase factors,
the order expansion of the three NP scalars $\phi_a,~a=0,1,2$ in powers of $r^{-1}$ fails to produce consistent and closed solutions---even when enlarging the solution space to include polyhomogeneous functions.

Further analysis based on the exact formal integrals (\ref{InteComplex}) shows that
the expansion in terms of LV coupling $k^2=k^t-\frac{k^r}{g(r)}$ is inherently nonperturbative.
This is because the natural length scale $1/|k^2|$ implies that for $r\gg1/|k^2|$,
the terms with higher power of $k^2$ is less suppressed compared to those of lower powers,
particularly as $r$ increases, making them non-negligible.
Although a naively suppression of the LV phase factor does not yield closed equations,
the expansion becomes well-defined and reduces to closed equations in the absence of LV,
and shows that the leading-order outgoing radiation mode $\phi_2\sim\mathcal{O}(r^{-1})$ remains unaltered.
Additionally, several expansion equations remain meaningful, such as the charge conservation Eq. (\ref{0th-u-Eq}),
which indicates that $\phi_2^0$ continues to serve as the news function.
In other words, given the time evolution of $\phi_2^0$, one can systematically solve for all the other terms order by order.

Astrophysical observations place very stringent constraints on $|k_{AF}|$ \cite{dataTable}\cite{CosBire2022} and, consequently, on $|k^2|$.
While the exact solutions indicate that the Coulomb mode $\phi_1$ can develop a $\mathcal{O}(\ln{r}/r^{2})$ behavior---deviating from the expected $\mathcal{O}(r^{-2})$ falloff
---the leading-order decay behaviors of $\phi_a$ effectively remain unaltered, namely $\phi_a^0\sim\mathcal{O}(r^{a-3}),~a=0,1,2$.
In other words, the peeling theorem remains valid at large but finite distances, at least up to $r\le6.4$pc based on a relatively conserved estimate.

From naive dimensional analysis, it is not entirely surprising that the decay behaviors of $\phi_a,~a=0,1,2$ may be altered at null infinity.
At the Lagrangian level, the operator $k^a{^*F}_{ab}A^b\supset2k^2(A^1\mathrm{Im}[\phi_1]+\mathrm{Im}[A^4\phi_2])$ contains one fewer derivative
compared to the LI operator $-\frac{1}{4}F^{\mn}F_{\mn}$.
Consequently, the dynamical equations derived from this Lagrangian exhibit a more moderate decay behavior as the radial distance $r$ increases.
Since the CPT-odd operator is a relevant operator, it has stronger influence at large distance than at short distance ---
meaning it primarily affects infrared than ultraviolet physics.
Based on this expectation, it is of no strange that $\phi_1$ develops a logarithmic correction, scaling as $\mathcal{O}(\ln{r}/r^{2})$,
while $\phi_0$ may behave as $\mathcal{O}(\ln{r}/r^{3})$.
However, these naive dimensional estimates require a more rigourous foundations,
such as a detailed analysis using a set of properly adapted ``null tetrad" suited for CPT-odd photons.
Furthermore, a closer scrutiny of the radiation flux at null infinity may provides further insight and is currently under development.

In addition, a detailed study of the large distance behavior of the $k_F$ operators remains an open question
and would be interesting in its own right.
While naive dimensional analysis suggests only quantitative deviation from the LI falloff behaviors,
there is no good reason to rule out unexpected surprises emerging from a more thorough investigation.

\section{Ackowledgement}
The author Z. Xiao would like to appreciate the valuable discussions with B. Altschul and M. Schreck, R. Lehnert and A. Kosteleck\'y
and also their selfless help and sharing of the very relevant references.
The authors want to express their gratitude for the great help of Bing Sun, Tao Zhu, Pujian Mao, Zhanfeng Mai and Dicong Liang.
This work is also supported by National Science Foundation of China under under grant No. 11605056.

\section{appendix}
\subsection{The determinant of the wave operator matrix}\label{detDRkF}
For the $k_F$-term modified Maxwell theory, the wave operator multiplying $A_\nu$ can be rewritten as
\bea&&
S^{\mn}=p^2[\Pi^{\mn}+\frac{1}{\xi}P^{\mn}+2\varkappa^{\mn}],
\qquad  P^{\mn}\equiv\frac{p^\mu p^\nu}{p^2},\,\nn&&
\Pi^{\mn}\equiv g^{\mn}-P^{\mn},\quad \varkappa^{\mn}\equiv(\bar{k}_F)^{\mu\rho\nu\si}P_{\rho\si}.
\eea
Then it is easy to verify that
\bea&&
\Pi^\mu_{~\rho}\Pi^{\rho}_{~\nu}=\Pi^{\mn},\quad P^\mu_{~\rho}P^{\rho}_{~\nu}=P^{\mn},\nn&&
\Pi^\mu_{~\rho}P^{\rho}_{~\nu}=\varkappa^{\mu}_{~\rho}P^{\rho}_{~\nu}=0,\quad
\varkappa^{\mu}_{~\rho}\Pi^{\rho}_{~\nu}=\varkappa^{\mn},
\eea
where in the last line, the equalities still hold true when left and right multipliers are interchanged,
such as $\Pi^{\mu}_{~\rho}\varkappa^{\rho}_{~\nu}=\varkappa^{\mn}$.
So by direct calculation, it is not difficulty to find that
\bea\label{n-powerS}&&
\hspace{-3.9mm}(S^n)^{\mu\nu}=(p^2)^n\left[\Pi^{\mn}+\frac{P^{\mn}}{\xi^n}+\sum_{m=1}^n\frac{2^m n!}{m!(n-m)!}(\varkappa^m)^{\mn}\right].
\nn&&
\eea
Then we can either use the formula in Ref.\cite{CLP2012} or by direct calculation to get the determinant of $(S^{\mn})$
up to the 2nd order of the LV coefficient as below
\bea&&\label{detSxi}
\hspace{-5mm}\mathrm{det}[S]=\frac{(p^2)^3}{\xi}\left\{p^2+2[\varkappa]p^2+2\varkappa^{\mn}p_\mu p_\nu(\xi-1)
\right.\nn&&\left.
+2p^2([\varkappa]^2-[\varkappa^2])+4(1-\xi)[(\varkappa^2)^{\mn}-[\varkappa]\varkappa^{\mn}]p_\mu p_\nu\right\}\nn&&
=\frac{(p^2)^3}{\xi}\left\{p^2+2([\varkappa]+[\varkappa]^2-[\varkappa^2])p^2\right\},
\eea
where the identities $(\varkappa^2)^{\mn}p_\mu p_\nu=\varkappa^{\mn}p_\mu p_\nu=0$ have been used and we denote the trace of a matrix by square bracket,
such as $[\varkappa]\equiv\mathrm{tr}[\varkappa]$.
Note $p^2=g_{\mn}p^\mu p^\nu$ can also be the scalar invariant of $p^\mu$ in curved spacetime.
From Eq. (\ref{detSxi}), we can find that two of the $p^2=0$ correspond to the gauge mode and the longitudinal mode, respectively,
and the remain equation reads
\bea&&\label{2ndDisRkF}
p^2[p^2+2(k_F)^{\mu\rho~\si}_{~~\mu}p_\rho p_\si]+2[(k_F)^{\mu\rho~\si}_{~~\mu}(k_F)^{\nu\al~\be}_{~~\nu}\nn&&
~~~~-(k_F)^{\mu\al~\be}_{~\,~\ga}(k_F)^{\ga\rho~\si}_{~~\mu}]p_\rho p_\si
p_\al p_\be=0.
\eea
This is a 4-th order equation and in general has 4 different solutions.
The symmetry means that for every solution $(\omega(\vec{p}), \vec{p})$, there is a corresponding solution $(-\omega(-\vec{p}), -\vec{p})$,
and we may interpret the later one as corresponding to anti-photon with negative energy.
Thus the two positive energy solutions corresponding to different polarizations,
and the two solutions are in general different and is the called vacuum birefringence.
Interestingly, the vacuum birefringence can occur even when only $c_F\neq0$ in $k_F$,
and the free of birefringence for the $c_F$ term is only meaningful at leading order \cite{Marco2015}.

\subsection{The differential forms and NP forms of Faraday tensor}\label{Null&DFs}
To get the Faraday tensor in terms of the NP scalars, it is better to take advantage of the simplicity of Maxwell's equation in
the Cartesian coordinates in Minkowski spacetime.
For simplicity, we assume the radiation propagates along the $z$-direction,
and the line elements of Minkowski spacetime implies the null 1-forms are
\bea&&\label{NTCartesian}
dl=\frac{1}{\sqrt{2}}[dt-dz], \quad d{n}=\frac{1}{\sqrt{2}}[dt+dz],\nn&&
d{m}=\frac{1}{\sqrt{2}}[dx+idy],\quad d{\bar{m}}=\frac{1}{\sqrt{2}}[dx-idy].
\eea
For simplicity, we may disregard the 1-form reminder ``$d$" and collectively denote the null 1-forms as
$\{F^a=F^a_{~\mu}dx^\mu,a=1,2,3,4\}=\{\bf{l},\bf{n},\bf{m},\bf{\bar{m}}\}$.
The dual null vectors read
\bea&&
\hspace{-6.mm}\prt_{l}=\frac{1}{\sqrt{2}}[\hat{e}_t-\hat{e}_z],\quad \prt_{n}=\frac{1}{\sqrt{2}}[\hat{e}_t+\hat{e}_z],\nn&&
\hspace{-6.mm}\prt_{m}=\frac{1}{\sqrt{2}}[\hat{e}_x-i\hat{e}_y],\quad \prt_{\bar{m}}=\frac{1}{\sqrt{2}}[\hat{e}_x+i\hat{e}_y].
\eea
By simply noting that $(\prt_{l})^\mu=-n^\mu=\eta^{\mn}n_\nu$, $(\prt_{n})^\mu=-l^\mu$, \etc,
we may also collectively denote the null vectors as $\{E_a=E_a^{~\mu}\prt_\mu,a=1,2,3,4\}=\{\prt_{l},\prt_{n},\prt_{m},\prt_{\bar{m}}\}=\{-\hat{n},-\hat{l},\hat{\bar{m}},\hat{m}\}$.
As mentioned in the main context, it is advantageous to use the 2-form instead of the Faraday tensor cause the former
is coordinate independent.
Reminding that the null 1-forms involve complex $\bf{m},\bf{\bar{m}}$,
it is instructive to use the 2-forms corresponding to the complex self-dual and anti-self-dual Faraday tensors
\bea&&\label{Dual&SelfD}
F_d=\hf(F-i^*F),\quad F_a=\hf(F+i^*F),\nn&&
^*F_d=iF_d,\quad\quad ^*F_a=-iF_a.
\eea
where in the second line we have used the fact that for any $p-$form $\Omega$,
$^{**}\Omega=(-1)^{p(N-p)+s}\Omega$ (where $s$ is the number of negative eigenvalues of the metric tensor)
and $N=\mathrm{dim}\mathcal{M}$.

By reversing Eq. (\ref{NTCartesian}) to express $\{dt,dx,dy,dz\}$ in terms of $\{F^a,a=1,2,3,4\}$,
the self-dual and anti-self-dual 2-forms are
\bea&&
\hspace{-4.5mm}F_d=\hf(F-i^*F)=
\hf E_+^idx^i\wedge dx^0-\frac{i}{4}\epsilon_{ijk}E_+^kdx^i\wedge dx^j\nn&&
=\hf E_+^z({\bf{n}\wedge\bf{l}+\bf{m}\wedge\bf{\bar{m}}})+\hf[(E_+^x+iE_+^y)\bf{\bar{m}}\wedge\bf{n}\nn&&
~~+(E_+^x-iE_+^y)]\bf{m}\wedge\bf{l}\nn&&
=\phi_1({\bf{l}\wedge\bf{n}-\bf{m}\wedge\bf{\bar{m}}})-\phi_2{\bf{l}\wedge\bf{m}}
+\phi_0{\bf{n}\wedge\bf{\bar{m}}}.\nn&&
\hspace{-4.5mm}F_a=\hf(F+i^*F)=
\hf E_-^idx^i\wedge dx^0
+\frac{i}{4}\epsilon_{ijk}E_-^kdx^i\wedge dx^j\nn&&
=\bar{\phi}_1({\bf{l}\wedge\bf{n}+\bf{m}\wedge\bf{\bar{m}}})-\bar{\phi}_2{\bf{l}\wedge\bf{\bar{m}}}
+\bar{\phi}_0{\bf{n}\wedge\bf{m}}.
\eea
where $E^i_{\pm}\equiv E^i\pm\,iB^i$ and $F_a=F_d^c$, the complex conjugate of $F_d$,
and we have also defined
\bea&&
\phi_0\equiv-\hf(E_+^x+iE_+^y)=F_{\mn}m^\mu l^\nu,\nn&&
\phi_1\equiv-\hf E_+^z=\hf F_{\mn}(n^\mu l^\nu+m^\mu\bar{m}^\nu),\nn&&
\phi_2\equiv\hf(E_+^x-iE_+^y)=F_{\mn}n^\mu\bar{m}^\nu,
\eea
according to the spin weight $s$ of each NP scalar, \ie, $\phi_a\mapsto e^{is\varphi}\phi_a,~a=0,1,2$ with $s=1-a$.
Spin weight is defined by the rotation phase properties of an object such as $\phi_1$ or ${\bf{n}\wedge\bf{m}}$ in the tangent plane of a sphere
spanned by $\bf{m},~\bf{\bar{m}}$. For example, a function constructed from $m^\mu$ such as $f_\mu m^\mu$ can have spin weight ``+1".

With $F_a,~F_d$ at hand, we can also reverse Eq. (\ref{Dual&SelfD}) to get the Faraday 2-form $F$ shown in Eq. (\ref{F-2formNull}), similarly for ${^*F}$.
Since differential forms are independent of the coordinate choice, we can also make sure that the final results of the above calculations are valid and independent of the choice we made
for our coordinates system and even hold true when spacetime is not flat.

\subsection{Energy momentum tensor (stress energy tensor)}\label{EMTqftvsgr}
As in the main context, we use $k$ to represent $k_{AF}$ for simplification.
The CPT-odd action is
\bea\label{IOddEM}&&
\hspace{-5mm}I_{O}=\hf\int d^4x\sqrt{-g}k_\ka\epsilon^{\ka\la\mn}A_\la F_{\mn}\nn&&
=-\int d^4x k_\ka\tilde{\epsilon}^{\ka\la\mn}A_\la\prt_{\mu}A_\nu,
\eea
where $\epsilon^{\ka\la\mn}\equiv-\frac{\tilde{\epsilon}^{\ka\la\mn}}{\sqrt{-g}}$ and the Levi-Civita symbol $\tilde{\epsilon}^{\ka\la\mn}$
takes values ``$\pm1$” depending on the even or odd permutations of 0123 for the upper indices, otherwise is simply 0.
The last equality is because $F_{\mn}=2\nabla_{[\mu}A_{\nu]}=2\prt_{[\mu}A_{\nu]}$ without torsion.
First we take the metric variation of $I_{O}$ to get the stress energy tensor, and we call it the metric gravity approach (MGA).
The result is
\bea\label{MAG-1}
T_O^{\mn}=\frac{2}{\sqrt{-g}}\frac{\delta I_O}{\delta g_{\mn}}=0.
\eea
This is because from the last equality of (\ref{IOddEM}), there is no explicit dependence of $I_O$ on $g_{\mn}$.
In other words, the Chern-Simons-Maxwell term does not directly couple to gravity.
Eq. (\ref{MAG-1}) can be confirmed by direct calculations.
Denote the Lagrangian density of $I_{O}$ as $\mathcal{L}_O=\hf k_\ka\epsilon^{\ka\la\mn}A_\la F_{\mn}$
and substitute it into the above definition of stress energy tensor, we get
\bea&&
\hspace{-6mm}T_O^{\mn}=g^{\mn}\mathcal{L}_O+2\frac{\delta \mathcal{L}_O}{\delta g_{\mn}}\nn&&
\hspace{-3mm}=g_{\mn} k_\ka{^*F}^{\ka\la}A_\la+k_\ka\frac{\delta \epsilon^{\ka\la\al\be}}{\delta g_{\mn}}A_\la F_{\al\be}\nn&&
\hspace{-3mm}=g^{\mn} k_\ka{^*F}^{\ka\la}A_\la+k_\ka\frac{g_{\mn}}{2(-g)^{-\hf}}\tilde{\epsilon}^{\ka\la\al\be}A_\la F_{\al\be}=0.
\eea
Next we derive the EMT with the field theory approach (FTA) by assuming that
{\small
\bea\label{FATstress}
\Th^{\mn}=g^{\mn}\mathcal{L}-\frac{\prt\mathcal{L}}{\prt(\nabla_\mu A_\rho)}\nabla^\nu A_\rho-\frac{i}{2}\nabla_\ka[K^{\ka\mn}-2K^{(\mu|\ka|\nu)}],
\eea
}\hspace{-0.5mm}where the last term in the square bracket is for symmetrization adopting the Belinfante procedure,
and $K^{\ka\mn}\equiv\frac{\prt\mathcal{L}}{\prt(\nabla_\ka A_\rho)}[\mathcal{I}^{\mn}]_\rho^{~\si} A_\si$ and
$K^{(\mu|\ka|\nu)}\equiv\hf[K^{(\mu\ka\nu)}+K^{(\nu\ka\mu)}]$.
We denote the EMT obtained from FTA by $\Th_{\mn}$ to distinguish it from the same object obtained from MGA, which is denoted by $T_{\mn}$.
Interestingly, applying this formula (\ref{FATstress}) to $\mathcal{L}_O$ gives
{\small
\bea
\Th^{\mn}_O=g^{\mn} k_\ka{^*F}^{\ka\la}A_\la-\epsilon^{\ka\la\mu\be}[k_\ka A_\la F^\nu_{~\be}+A^\nu\nabla_\be(k_\ka A_\la)],
\eea
}\hspace{-0.5mm}which is neither symmetric nor zero.
The conflict between these two approaches may be solved by noting that 

Next we apply the two different approaches to the CPT-even $k_{F}$ term. For completeness, we also include the LI Maxwell actions, \ie,
\bea\label{IEvenEM}&&
\hspace{-5mm}I_E=-\frac{1}{4}\int d^4x\sqrt{-g}\chi^{\al\be\ga\rho}F_{\al\be}F_{\ga\rho},
\eea
where $\chi^{\al\be\ga\rho}\equiv\,g^{\al[\ga}g^{\rho]\be}+(k_F)^{\al\be\ga\rho}$. The MGA gives
\bea\label{MGAstressE}
T^{\mn}_E=[F^{\mu\al}F^{\nu}_{~\al}+2F^{(\nu}_{~~\al}(k_F)^{\mu)\al\be\ga}F_{\be\ga}]+g^{\mn}\mathcal{L}_E.
\eea
The LI part is unaltered and we focus our discussion on the $k_F$ term.
We need to keep in mind that the upper indices $k^\Xi$ (where $k^\Xi$ denotes a generic background observer tensor field and $\Xi$ denotes Lorentz indices)
can be physically quite distinct from the lower indices $k_\Xi$ unless they are the background fields which trigger spontaneously Local Lorentz symmetry breaking \cite{ZKbkgGravity}.
In this context, we adopt the lower indices convention for all the constant or background fields, such as the Levi-Civita symbol and $(k_F)_{\al\be\ga\rho}$.
Then the metrics are all encoded in the upper indices dynamical fields and $\sqrt{-g}$.
For example, $\delta F^{\rho\si}=F_{\mn}\delta(g^{\rho\mu}g^{\si\nu})=-\delta g_{\mn}[F^{\rho\mu}g^{\si\nu}+F^{\mu\si}g^{\rho\nu}]$.
Then we find
\bea&&
\delta L_{EV}
=\hf\delta g_{\mn}g^{\mn}L_{EV}-\frac{1}{4}\sqrt{-g}(k_F)_{\al\be\ga\rho}2F^{\al\be}\delta F^{\ga\rho}\nn&&
\hspace{-5mm}=\hf\{g^{\mn}L_{EV}+2\sqrt{-g}F^{(\mu}_{~~\rho}(k_F)^{\nu)\rho\al\be}F_{\al\be}\}\delta g_{\mn},
\eea
where $L_{EV}\equiv-\frac{1}{4}\sqrt{-g}(k_F)_{\al\be\ga\rho}F^{\al\be}F^{\ga\rho}$ and at the last step, we intentionally symmetrize the $k_F$ term by assuming $g_{\mn}=g_{(\mn)}$,
though this is not necessarily true unless with the test particle assumption.
Substituting into $T^{\mn}_E=\frac{2\delta I_{E}}{\sqrt{-g}\delta g_{\mn}}$ gives Eq. (\ref{MGAstressE}).
Note the test particle assumption give rise to inconsistency if compare the symmetric $T^{\mn}_E$ obtained by assuming
a symmetric $g_{(\mn)}$ with the apparently asymmetric $\Th^{\mn}_E$ obtained below.

Substituting $\mathcal{L}_E=-\frac{1}{4}\chi_{\al\be\ga\rho}F^{\al\be}F^{\ga\rho}$ into Eq. (\ref{FATstress}), we can derive the stress energy tensor,
{\small
\bea
\Th^{\mn}_E=\chi^{\al\be\mu\rho}F^{\nu}_{~\rho}F_{\al\be}+g^{\mn}\mathcal{L}_{E}
-A^\nu\nabla_\rho[\chi^{\al\be\mu\rho}F_{\al\be}].
\eea
}\hspace{-0.6mm}Note the immediate results in deriving $\Th^{\mn}_E$ from the FTA are
{\small
\bea
\frac{\prt\mathcal{L}_E}{\prt(\nabla_\mu A_\rho)}=-\chi^{\al\be\mu\rho}F_{\al\be},
~
K_3^{\mu\ka\nu}=2i\,\chi^{\ka\mu\al\be}F_{\al\be}A^\nu.
\eea
}

\end{document}